\documentclass[%
 reprint,
 amsmath,amssymb,
 aps,
]{revtex4-2}

\usepackage{graphicx}
\usepackage{dcolumn}
\usepackage{bm}

\usepackage{amsmath}   
\usepackage{amssymb}   
\usepackage{braket}    
\usepackage{quantikz}
\usepackage{tikz}
\usetikzlibrary{quantikz2}
\usepackage{adjustbox}
\usepackage{array}
\usepackage{braket}
\usepackage{float}
\usepackage{subcaption}
\usepackage{graphicx}
\usepackage{hyperref}
\usepackage{enumitem}
\usepackage{mathtools}
\mathtoolsset{showonlyrefs}
\usepackage{xcolor}

\begin{document}

\title{Gate-based emulation of boson sampling using photonic qubits}

\author{Aastha P. Zalone}
\affiliation{%
Quantum Optics \& Quantum Information, Department of Electronic Systems Engineering, Indian Institute of Science, Bengaluru 560012, India
}
\author{S. P. Dinesh}%
\affiliation{%
Quantum Optics \& Quantum Information, Department of Electronic Systems Engineering, Indian Institute of Science, Bengaluru 560012, India
}

\author{C. M. Chandrashekar}%
\affiliation{%
Quantum Optics \& Quantum Information, Department of Electronic Systems Engineering, Indian Institute of Science, Bengaluru 560012, India
}


\begin{abstract}
Boson sampling arising from multiphoton interference in linear-optical networks is a prominent non-universal model for quantum computation. Here, by encoding the multi-qubit state to bosonic Fock state, we present a scalable quantum-circuit framework for simulating boson sampling on a universal quantum computing platform. Beginning with balanced beam-splitter transformations on the single- and two-photon sectors, we derive equivalent quantum-circuit implementations and unify them within a common Hilbert-space representation using an ancilla-assisted encoding. This construction is then generalized to arbitrary interferometers by replacing each optical beam splitter with a repeating quantum-circuit unit that selectively acts only within the relevant local interference subspace, requiring $N+1$ qubits for a two-photon $N$-mode interferometer and a linear-overhead subspace-identification procedure. Using this framework, gate-based quantum circuit for a four-mode boson-sampling circuit is developed and experimentally implemented on a four-qubit gate-based photonic qubit system. The qubit framework for emulating boson sampling of $n-$photons in $m-$mode will be useful to solve a broad class of sampling complexity problem on a gate-based quantum computers.
\end{abstract}


\maketitle



\section{Introduction}
\label{Intro}


Quantum computation offers the potential to solve certain problems that are computationally intractable for classical computers~\cite{feynman1982,deutsch1985}. However, the realization of large-scale universal quantum computers remains challenging due to noise, decoherence, and limitations in scalability. These challenges have motivated the exploration of restricted models of quantum computation that are experimentally accessible while retaining the potential for computational advantage. A prominent example among them is boson sampling~\cite{brod2019photonic} , a non-universal model of quantum computation based on the interference of multiple indistinguishable photons in linear-optical networks.

Originally proposed by Aaronson and Arkhipov~\cite{Aaronson2011}, boson sampling considers multiple indistinguishable photons injected into a linear-optical interferometer described by a unitary transformation~\cite{reck1994,clements2016}. Repeated measurements of the output photon occupations generate samples from the corresponding multiphoton probability distribution. The transition amplitudes between input and output Fock states are determined by permanents of submatrices of the interferometer unitary, whose exact evaluation is \#P-hard for general matrices~\cite{Valiant1979}. This connection, together with additional complexity-theoretic arguments, provides the basis for the expected classical intractability of boson sampling at sufficiently large scales~\cite{Aaronson2011}. Several variants have subsequently been introduced to address different theoretical and experimental settings, including scattershot boson sampling using heralded probabilistic sources~\cite{lund2014}, Gaussian boson sampling employing squeezed-state inputs~\cite{hamilton2017}, and lossy boson sampling accounting for experimentally unavoidable photon loss~\cite{wang2018loss}. Methods for characterizing and validating boson-sampling output statistics have also been developed, including efficient tests for distinguishing boson-sampling distributions from uniform sampling~\cite{aaronson2014far}.

Following its theoretical proposal, boson sampling was experimentally demonstrated using both bulk and integrated linear-optical platforms~\cite{spring2013boson,tillmann2013experimental,broome2013photonic}. Integrated multimode interferometers were subsequently developed as compact and configurable linear-optical networks for implementing photonic boson sampling~\cite{crespi2013integrated}. Further experiments introduced methods for validating boson-sampling devices and implemented scattershot architectures to increase multiphoton sampling rates and improve scalability~\cite{spagnolo2014experimental,bentivegna2015experimental}. More recently, large-scale Gaussian boson-sampling experiments have reported quantum computational advantage, demonstrating the ability of photonic sampling architectures to access regimes that are challenging to simulate using classical computers~\cite{zhong2020quantum,madsen2022quantum}. These advances highlight the suitability of photonic systems for quantum information processing, supported by multiple degrees of freedom for encoding quantum information, coherent linear-optical control, and increasingly advanced single-photon generation and detection technologies~\cite{flamini2019photonic}.

Bosonic and linear-optical dynamics have also been investigated within the gate-based quantum-computing framework. Several qubit encodings of truncated bosonic Hilbert spaces, including unary, binary, Gray-code, and compact representations, have been developed to represent bosonic states on quantum registers~\cite{sawaya2020,kirby2021}. Gate-based simulations of linear-optical interference phenomena, including Hong--Ou--Mandel interference, have also been demonstrated on quantum processors~\cite{mohan2025}, while quantum-circuit formulations of boson sampling have been developed to incorporate effects such as partial photon distinguishability~\cite{moylett2018quantum}. Related bosonic dynamics and sampling protocols have further been proposed or implemented on non-photonic quantum platforms, including trapped-ion and superconducting-qubit systems~\cite{shen2014,lamata2014,li2026}. Together, these developments demonstrate the broader applicability of representing boson sampling circuit within qubit- and gate-based quantum-computing frameworks.

Despite these developments, boson sampling is most naturally formulated in the Fock basis, where linear-optical transformations act directly on photon-number states, whereas gate-based quantum computers operate on qubit registers through sequences of elementary quantum gates. Bridging these descriptions therefore requires an efficient encoding of the relevant bosonic Fock states into qubits together with a circuit representation of the corresponding linear-optical transformations. In particular, mapping bosonic Fock-space dynamics onto qubit operations provides a hardware-agnostic logical description, making the resulting circuits, in principle, independent of the physical platform used to realize the qubits.

In this work, we develop a quantum-circuit framework for representing linear-optical beam-splitter transformations~\cite{knill2001} within a qubit encoding of bosonic Hilbert spaces. We begin by constructing quantum circuits for the single-photon and two-photon sectors of a balanced beam splitter that reproduce the corresponding Fock-space transformations. These constructions are then combined through a unified encoding that accommodates both photon-number sectors within a common qubit representation. Using the resulting beam-splitter circuit as a fundamental building block, we extend the construction to multimode linear-optical networks through a scalable repeating-unit architecture. The resulting framework provides a systematic procedure for translating boson-sampling transformations from their conventional Fock-space description into gate-based quantum circuits.  We experimentally validate the framework by implementing the gate-based quantum circuit for one- and two-photon four-mode boson-sampling on a four-qubit photonic qubits system.

This work is reported in the following structure.  In  Sec.\,\ref{Thbk} we will introduce to the basic description of the linear optical network, boson sampling model and beam splitter transformation that will serve as a background for the remaining part of the work. Sec.\,\ref{sec:level1} presents the mapping for the Fock state with the computation basis and quantum circuit representation of the beam splitter operators on single, one- and two-photon input modes.  In Sec.\,\ref{sec:generalised_2_photon} quantum circuit generalization for two-photon N-mode boson sampling and in Sec.\,\ref{sec:four_mode} an explicit quantum circuit on a four-qubit system is presented. In Sec.\,\ref{Expt} an experimental realization outlining the four-qubit photonic state generation and implementation of a gate-based quantum circuit for boson sampling along with the experimental result is reported. Section\,\ref{Conc} summarizes the results and discusses their potential impact in various fields. Finally, for the sake of readability, we group several appendices with the explicit calculations for the interferometer transformation, the effective models, and the boson-sampling state preparation.

\section{Theoretical Background}
\label{Thbk}

\subsection{Linear Optical Networks}

A passive linear optical interferometer consisting of beam splitters and phase shifters is completely characterized by a unitary matrix $U \in U(m)$, where $m$ denotes the number of optical modes \cite{reck1994, clements2016}. The creation operators transform according to
\begin{equation}
\hat{b}_i^\dagger
=
\sum_{j=1}^{m}
U_{ij}\hat{a}_j^\dagger ,
\label{eq:linear}
\end{equation}
where $\hat{a}_j^\dagger$ and $\hat{b}_i^\dagger$ denote the input and output mode creation operators, respectively. Since passive optical elements preserve the total photon number, the evolution of an arbitrary photonic state is completely determined by the interferometer unitary $U$.

\subsection{Boson Sampling}

Boson sampling considers the propagation of indistinguishable photons through a passive linear optical network followed by photon-number measurements at the output. Perfect photon indistinguishability is essential for the many-particle interference underlying ideal boson sampling, while partial
distinguishability modifies the resulting output probability distribution \cite{tichy2015sampling}. The input and output configurations are described in the Fock basis
\begin{equation}
\ket{n_1,n_2,\ldots,n_m},
\label{eq:fock}
\end{equation}
where $n_i$ denotes the occupation number of mode $i$.
For an input occupation vector
\begin{equation}
\mathbf{n}
=
(n_1,n_2,\ldots,n_m),
\end{equation}
and an output occupation vector
\begin{equation}
\mathbf{s}
=
(s_1,s_2,\ldots,s_m),
\end{equation}
the transition probability is
\begin{equation}
P(\mathbf{n}\rightarrow\mathbf{s})
=
\left|
\braket{\mathbf{s}|U|\mathbf{n}}
\right|^{2}.
\label{eq:prob}
\end{equation}
The corresponding transition amplitude is given by 
\begin{equation}
\braket{\mathbf{s}|U|\mathbf{n}}
=
\frac{
\mathrm{Perm}\!\left(U_{\mathbf{s},\mathbf{n}}\right)
}
{
\sqrt{\prod_i n_i!\prod_j s_j!}
},
\label{eq:amp}
\end{equation}
where $U_{\mathbf{s},\mathbf{n}}$ denotes the submatrix of $U$ specified by the input and output occupations\,\cite{Aaronson2011} \cite{Troyansky1996}\cite{scheel2008}. The matrix permanent is defined as
\begin{equation}
\mathrm{Perm}(A)
=
\sum_{\sigma \in S_N}
\prod_{i=1}^{N}
A_{i,\sigma(i)},
\label{eq:perm}
\end{equation}
where $S_N$ denotes the set of all permutations of $N$ elements\,\cite{Valiant1979}.

The computational complexity associated with evaluating matrix permanents provides strong evidence for the classical hardness of large-scale boson sampling\,\cite{Valiant1979,Aaronson2011}.

\subsection{Beam Splitter Transformation}

The beam splitter forms the fundamental building block of passive linear optical networks. For a balanced $50{:}50$ beam splitter, the corresponding unitary transformation is
\begin{equation}
U_{\mathrm{BS}}
=
\frac{1}{\sqrt{2}}
\begin{pmatrix}
1 & 1 \\
1 & -1
\end{pmatrix}.
\label{eq:BS}
\end{equation}
The associated transformation of the mode operators is
\begin{align}
\hat{a}^{\dagger}
&\rightarrow
\frac{\hat{a}^{\dagger}+\hat{b}^{\dagger}}
{\sqrt{2}},
\\
\hat{b}^{\dagger}
&\rightarrow
\frac{\hat{a}^{\dagger}-\hat{b}^{\dagger}}
{\sqrt{2}}.
\label{eq:BSop}
\end{align}

Application of Eq.~(\eqref{eq:BSop}) to multiphoton Fock states gives rise to non-classical interference effects arising from bosonic indistinguishability. In particular, the interference of two identical photons at a balanced beam splitter leads to the Hong--Ou--Mandel effect ~\cite{Hong1987} , in which coincidence events are suppressed due to destructive interference between indistinguishable photon paths. In the following section, these beam splitter transformations are mapped onto quantum-circuit representations within bosonic Hilbert space.

\section{\label{sec:level1}QUANTUM-CIRCUIT REPRESENTATION OF THE BEAM SPLITTER}
\subsection{Case I: Single-Photon Input}

We first consider the action of a balanced beam splitter in the single-photon sector spanned by the Fock states
\[
\{|1,0\rangle, |0,1\rangle\}.
\]
Under a balanced $50{:}50$ beam splitter transformation,
\[
\begin{aligned}
|1,0\rangle &\rightarrow
\frac{|1,0\rangle + |0,1\rangle}{\sqrt{2}}, \\
|0,1\rangle &\rightarrow
\frac{|1,0\rangle - |0,1\rangle}{\sqrt{2}}.
\end{aligned}
\]

To realize this transformation on a gate-based quantum computer, each optical mode is mapped directly onto a qubit \cite{Nielsen2010}. Since this state space is two-dimensional, the transformation requires only one qubit. However, to maintain a consistent mode-to-qubit encoding that facilitates generalization to larger interferometers, to be explained in Sec. \ref{sec:generalised_2_photon}, we embed the single-photon state space into a two-qubit computational basis. The associated circuit implementation is shown in Fig.\ref{fig:1 photon circuit}, while the explicit matrix representation is provided in the Appendix \ref{sec: BS_1_photon_transformation}.

\begin{table}[h]
\centering
\caption{Encoding for the single-photon sector.}
\label{tab:singlephoton}
\begin{tabular}{cc}
\hline
Fock State & Computational Basis State \\
\hline
$\ket{1,0}$ & $\ket{10}$ \\
$\ket{0,1}$ & $\ket{01}$ \\
\hline
\end{tabular}
\end{table}
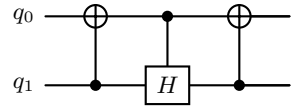
\begin{figure}[htpb]
    \centering
    \begin{quantikz}
        \lstick{$q_0$} & \targ{} & \ctrl{1} & \targ{} & \qw \\
\lstick{$q_1$} & \ctrl{-1} & \gate{H} & \ctrl{-1} & \qw
    \end{quantikz}
    \caption{Quantum circuit representation of the beam splitter transformation in the single-photon sector under the two-qubit embedding.}
    \label{fig:1 photon circuit}
\end{figure}

\subsection{Case II: Two-Photon Input}
\label{sec3b}

We now consider the action of a balanced beam splitter in the two-photon sector spanned by the Fock states
\[
\{|2,0\rangle, |0,2\rangle, |1,1\rangle\}.
\]
The two-photon input states is embedded into the two-qubit computational basis according to Table~\ref{tab:two_photon_encoding}.
\begin{table}[h]
\centering
\caption{Binary occupation encoding for the two-photon sector.}
\label{tab:two_photon_encoding}
\begin{tabular}{cc}
\hline
Fock State & Computational Basis State \\
\hline
$|0,0\rangle$ & $|00\rangle$ \\
$|0,2\rangle$ & $|01\rangle$ \\
$|2,0\rangle$ & $|10\rangle$ \\
$|1,1\rangle$ & $|11\rangle$ \\
\hline
\end{tabular}
\end{table}
For a balanced $50{:}50$ beam splitter, the two-photon states evolve according to
\[
\begin{aligned}
|2,0\rangle \rightarrow \;&
\frac{1}{2}|2,0\rangle
+
\frac{1}{2}|0,2\rangle
+
\frac{1}{\sqrt{2}}|1,1\rangle,
\\[6pt]
|0,2\rangle \rightarrow \;&
\frac{1}{2}|2,0\rangle
+
\frac{1}{2}|0,2\rangle
-
\frac{1}{\sqrt{2}}|1,1\rangle,
\\[6pt]
|1,1\rangle \rightarrow \;&
\frac{1}{\sqrt{2}}|2,0\rangle
-
\frac{1}{\sqrt{2}}|0,2\rangle.
\end{aligned}
\]
The transformation exhibits Hong--Ou--Mandel interference, where the coincidence component is suppressed due to bosonic bunching at the beam splitter. The associated circuit decomposition is shown in Fig.\ref{fig:circuits}, while the explicit matrix representation is provided in the Appendix \ref{sec: BS_2_photon_transformation}.
\begin{figure}[htpb]
    \centering

    \begin{quantikz}
        \lstick{$q_0$} & \targ{} & \ctrl{1} & \targ{} & \ctrl{1} & \targ{} & \\
        \lstick{$q_1$} & \ctrl{-1} & \gate{H} & \ctrl{-1} & \gate{H} & \ctrl{-1} & 
    \end{quantikz}

    \caption{Quantum circuit implementations of the encoded beam splitter unitary $U_{\text{Fock}}$.}
    \label{fig:circuits}
\end{figure}
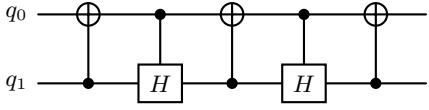

\subsection{Case III: One- and Two-Photon Inputs}
\label{sec3c}

The single-photon and two-photon sectors discussed above can be combined into a unified Hilbert space containing all states with photon number less than or equal to two. Such a representation is useful for constructing scalable quantum circuits, since it allows vacuum, single-photon, and two-photon states to be treated within a common computational basis.

For two spatial modes, the Fock basis is
\begin{equation}
\mathcal{B}_{\le 2}
=
\{
|2,0\rangle,
|0,2\rangle,
|1,1\rangle,
|1,0\rangle,
|0,1\rangle,
|0,0\rangle
\}.
\end{equation}
To embed this six-dimensional Hilbert space into a qubit register, we employ a three-qubit encoding. The first two qubits retain the encoding used in the previous sections, while the third qubit distinguishes between the single-photon and two-photon sectors.

This choice is motivated by the circuit structure obtained in the previous sections. The quantum circuits corresponding to the single-photon and two-photon beam splitter transformations share the same first three gates. The difference arises only in the final two gates, which are required exclusively to reproduce the non-classical two-photon interference. Consequently, these final two gates should act only on the two-photon subspace, while leaving the single-photon states unaffected.

To achieve this selective application, an additional qubit is introduced to identify whether the encoded state belongs to the two-photon sector. The final two gates are then conditioned on this qubit, allowing the circuit to reproduce both the single-photon and two-photon beam splitter dynamics within a unified one- and two-photon Hilbert space. The encoding is defined in Table \ref{tab:binary_occupation_encoding}.
\begin{table}[h]
\centering
\caption{Encoding for vacuum, single and two photon input}
\label{tab:binary_occupation_encoding}
\begin{tabular}{cc}
\hline
Fock State & Computational Basis State \\
\hline
$|0,0\rangle$ & $|000\rangle$ \\
$|0,1\rangle$ & $|010\rangle$ \\
$|1,0\rangle$ & $|100\rangle$ \\
$|1,1\rangle$ & $|111\rangle$ \\
$|0,2\rangle$ & $|011\rangle$ \\
$|2,0\rangle$ & $|101\rangle$ \\
\hline
\end{tabular}
\end{table}

Under this encoding, states belonging to the two-photon sector are characterized by the third qubit being in the state $\ket{1}$, whereas the vacuum and single-photon states correspond to $\ket{0}$ in the third qubit.

\begin{figure}[htpb]
    \centering
    \begin{quantikz}
        \lstick{$q_0$} & \targ{}   & \ctrl{1} & \targ{}   & \ctrl{1}  & \targ{}    & \\
        \lstick{$q_1$} & \ctrl{-1} & \gate{H} & \ctrl{-1} & \gate{H}  & \control{} & \\
        \lstick{$q_2$} &           &          &           & \ctrl{-1} & \ctrl{-2}  & 
    \end{quantikz}
    \caption{Quantum circuit implementing the beam splitter transformation in the Hilbert space with maximum photon number equal to two.}
    \label{fig:max_two_photon_circuit}
\end{figure}
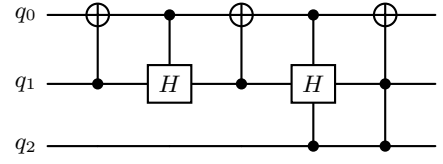

The corresponding quantum circuit is shown in Fig.\,\ref{fig:max_two_photon_circuit}. This circuit serves as the repeating unit of the generalized architecture developed in the following section.

\section{GENERALIZED TWO-PHOTON N-MODE QUANTUM CIRCUIT FOR BOSON SAMPLING}
\label{sec:generalised_2_photon}

The unified beam-splitter circuit developed in Sec.\,\ref{sec3c} forms the fundamental building block of the generalized construction presented in this section. We now extend this framework to the two-photon boson-sampling model in an arbitrary number of optical modes. The central idea is to replace each beam splitter appearing in a linear-optical interferometer by an equivalent quantum-circuit block acting on an encoded qubit register. By concatenating these blocks according to the structure of the interferometer, a complete quantum-circuit representation of the boson-sampling dynamics can be obtained.

\subsection{Encoding of Two-Photon States}

Consider two photons distributed among $N$ spatial modes. The corresponding Fock states belong to one of two classes:
\begin{enumerate}
    \item Both photons occupying the same mode,
    \begin{equation}
    |0,\ldots,2_k,\ldots,0\rangle ,
    \end{equation}
    for $k=1,\ldots,N$.
        \item Photons occupying two distinct modes,
    \begin{equation}
    |0,\ldots,1_i,\ldots,1_j,\ldots,0\rangle ,
    \qquad i\neq j .
    \end{equation}
\end{enumerate}

To encode these states, one qubit is assigned to each optical mode. The computational basis state
\begin{equation}
|q_1 q_2 \ldots q_N\rangle
\end{equation}
records the set of occupied modes, where
\begin{equation}
q_i=
\begin{cases}
1, & \text{if mode } i \text{ is occupied},\\
0, & \text{otherwise}.
\end{cases}
\end{equation}
Since the construction is restricted to the two-photon sector, it is sufficient to encode only the occupied modes. Consequently, a computational basis state containing a single logical $1$ represents a configuration in which both photons occupy the corresponding mode, whereas a basis state containing two logical $1$'s represents one photon in each occupied mode. Consequently, the number of data qubits required is
\begin{equation}
N_q=N.
\end{equation}

An additional ancilla qubit is introduced to identify the local interference subspace, giving a total register size of
\begin{equation}
N+1
\end{equation}
qubits.

\subsection{Identification of the Interference Subspace}
Consider a beam splitter acting on modes $i$ and $j$. As shown in Sec.\,\ref{sec3b}, in the two-photon sector, non-trivial interference at a beam splitter $(i,j)$ 
is restricted to the Fock states 
\begin{equation}
\{|2_i0_j\rangle,\;|0_i2_j\rangle,\;|1_i1_j\rangle\}.
\end{equation}

All modes other than i and j are referred to as spectator modes.
For two-photon interference to occur at beam splitter $(i,j)$, both photons must occupy modes $i$ and/or $j$. Consequently, all remaining modes act as spectator modes and must be unoccupied. Hence required condition for a valid two photon interference is
\begin{equation}
q_k=0,
\qquad \forall k\neq i,j.
\end{equation}

This condition is encoded into the ancilla according to
\begin{equation}
a=
\bigwedge_{k\neq i,j}\neg q_k,
\label{eq:global_logical_condition}
\end{equation}

Anti-controlled CNOT and anti-controlled Toffoli gates are applied with the spectator qubits as controls and the ancilla as the target. Since the multi-qubit AND condition cannot be implemented directly as a single elementary gate, it is decomposed into a sequence of two- and three-qubit gates. By applying these gates successively across the spectator qubits, the global logical condition (Eq. \ref{eq:global_logical_condition}) is realized within the circuit. The ancilla is therefore activated if and only if the encoded state belongs to the local interference subspace. 


\subsection{Resource Scaling}
For a beam splitter acting on modes $i$ and $j$, the number of spectator modes is
\begin{equation}
N_s=N-2.
\end{equation}
The ancilla-identification circuit is implemented using a sequence of anti-controlled CNOT and Toffoli gates. The required number of Toffoli gates is
\begin{equation}
N_{\mathrm{Toffoli}}
=
\left\lfloor
\frac{N-2}{2}
\right\rfloor ,
\end{equation}
where $\lfloor\cdot\rfloor$ denotes the floor function.

The required number of CNOT gates is
\begin{equation}
N_{\mathrm{CNOT}}
=
\begin{cases}
1, & N \text{ odd and } N\ge5,\\
0, & \text{otherwise}.
\end{cases}
\end{equation}
Hence, the total number of gates required to identify the interference subspace is
\begin{equation}
G=
\left\lfloor
\frac{N-2}{2}
\right\rfloor
+
\mathbf{1}_{\{N\ \mathrm{odd},\,N\ge5\}},
\end{equation}
where $\mathbf{1}$ denotes the indicator function.

The identification cost therefore scales linearly with the number of optical modes,
\begin{equation}
G = O(N),
\end{equation}
demonstrating that the proposed architecture remains scalable for large boson-sampling interferometers.

The ancilla initialization depends on the parity of the identification circuit. Let $G$ denote the total number of anti-controlled CNOT and Toffoli gates. The ancilla is initialized in $|1\rangle$ for even $G$ and in $|0\rangle$ for odd $G$, ensuring that it is activated exclusively for states belonging to the local interference subspace. After applying the repeating unit, the ancilla is uncomputed (discarded).
\begin{table}[h]
\centering
\begin{tabular}{c c c c c}
\hline
Fock states & $q_i$ & $q_j$ & $a$ (after Toffoli and CNOT) & Encoded state \\
\hline
$\ket{2,0}$ & 1 & 0 & 1 & 101 \\
$\ket{0,2}$ & 0 & 1 & 1 & 011 \\
$\ket{1,1}$ & 1 & 1 & 1 & 111 \\
$\ket{1,0}$ & 1 & 0 & 0 & 100 \\
$\ket{0,1}$ & 0 & 1 & 0 & 010 \\
$\ket{0,0}$ & 0 & 0 & 0 & 000 \\
\hline
\end{tabular}
\caption{Encoding of truncated two-photon Fock states into qubit basis.}
\label{tab:fock_encoding}
\end{table}

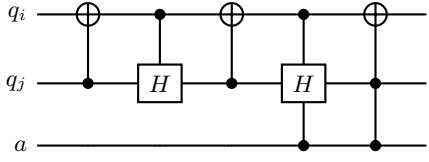
\begin{figure}[htpb]
    \centering
    \begin{quantikz}
        \lstick{$q_i$} & \targ{}   & \ctrl{1} & \targ{}   & \ctrl{1}  & \targ{}   & \\
        \lstick{$q_j$} & \ctrl{-1} & \gate{H} & \ctrl{-1} & \gate{H}  & \ctrl{-1} & \\
        \lstick{$a$}   &           &          &           & \ctrl{-1} & \ctrl{-1} & 
    \end{quantikz}
    \caption{Repeating unit of the generalized two-photon beam splitter circuit.}
   \label{fig:repeating_unit}
\end{figure}

Once the ancilla identifies a valid interference configuration, the repeating unit shown in Fig.~\ref{fig:repeating_unit} is applied to $(q_i,q_j,a)$. The ancilla-controlled gates implement the two-photon beam-splitter transformation derived in Sec.\,\ref{sec:level1} while leaving all other computational basis states unchanged. The ancilla is subsequently uncomputed by reversing the identification circuit. This repeating unit serves as the elementary building block of the generalized architecture.

\subsection{Construction of Arbitrary Boson-Sampling Circuits}
Any passive linear-optical interferometer can be decomposed into a sequence of beam splitters and phase shifters \cite{reck1994,clements2016}. Within the proposed framework, each beam splitter is replaced by the repeating unit described above, which acts locally on the corresponding pair of modes and a single ancilla. The complete gate-based circuit is thus constructed by sequentially applying the repeating unit according to the beam-splitter decomposition of the optical network. This provides a systematic mapping from a two-photon, N-mode linear-optical interferometer to a qubit-based quantum circuit requiring N+1 qubits, with a linear-overhead subspace-identification procedure. Since the repeating unit is independent of the total number of modes, the construction naturally extends to interferometers of arbitrary size while preserving the underlying two-photon interference structure.

\section{Two-Photon Four-Mode Boson-Sampling Circuit}
\label{sec:four_mode}
\begin{figure}[H]
    \centering
    \includegraphics[width=0.5\linewidth]{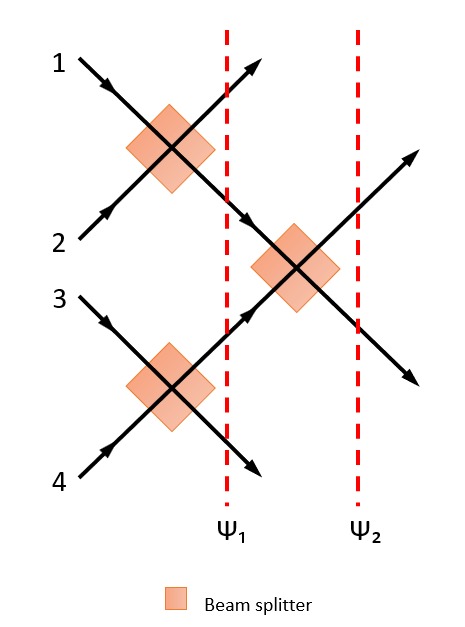}
    \caption{Schematic representation of the four-mode linear optical interferometer composed of three balanced beam splitters. The first interferometric stage consists of beam splitters $B_{12}$ and $B_{34}$ acting on modes $(1,2)$ and $(3,4)$, respectively, producing the intermediate state $\ket{\psi_{1}}$. The second stage comprises beam splitter $B_{23}$ acting on the central modes $(2,3)$, resulting in the final output state $\ket{\psi_{2}}$.}
    \label{fig:four_mode_linear_optical_interferometer}
\end{figure}

We now apply the generalized construction to two photons propagating through the four-mode interferometer shown in Fig.~\ref{fig:four_mode_linear_optical_interferometer}. The network consists of balanced beam splitters acting on the mode pairs $(1,2)$ and $(3,4)$, followed by a beam splitter acting on the central modes $(2,3)$. The corresponding single-particle transformation is denoted by $U_{4}$.

The two-photon Hilbert space has dimension
\begin{equation}
\dim \mathcal{H}^{(2)}_{4}
=
\binom{4+2-1}{2}
=
10,
\label{eq:four_mode_dimension}
\end{equation}
with Fock basis
\begin{equation}
\begin{split}
\mathcal{B}^{(2)}_{4}
=
\{&
\ket{2000},
\ket{0200},
\ket{0020},
\ket{0002},\\
&
\ket{1100},
\ket{1010},
\ket{1001},
\ket{0110},
\ket{0101},
\ket{0011}
\}.
\end{split}
\label{eq:four_mode_basis}
\end{equation}
Following the encoding introduced in the previous section, one qubit is associated with each optical mode. The four bunched states are encoded as
\begin{equation}
\begin{aligned}
\ket{2000} &\longrightarrow \ket{1000}, &
\ket{0200} &\longrightarrow \ket{0100},\\
\ket{0020} &\longrightarrow \ket{0010}, &
\ket{0002} &\longrightarrow \ket{0001},
\end{aligned}
\label{eq:bunched_encoding}
\end{equation}
while the six input states retain their binary occupation patterns,
\begin{equation}
\begin{aligned}
\ket{1100} &\longrightarrow \ket{1100}, &
\ket{1010} &\longrightarrow \ket{1010},\\
\ket{1001} &\longrightarrow \ket{1001}, &
\ket{0110} &\longrightarrow \ket{0110},\\
\ket{0101} &\longrightarrow \ket{0101}, &
\ket{0011} &\longrightarrow \ket{0011}.
\end{aligned}
\label{eq:collision_free_encoding}
\end{equation}

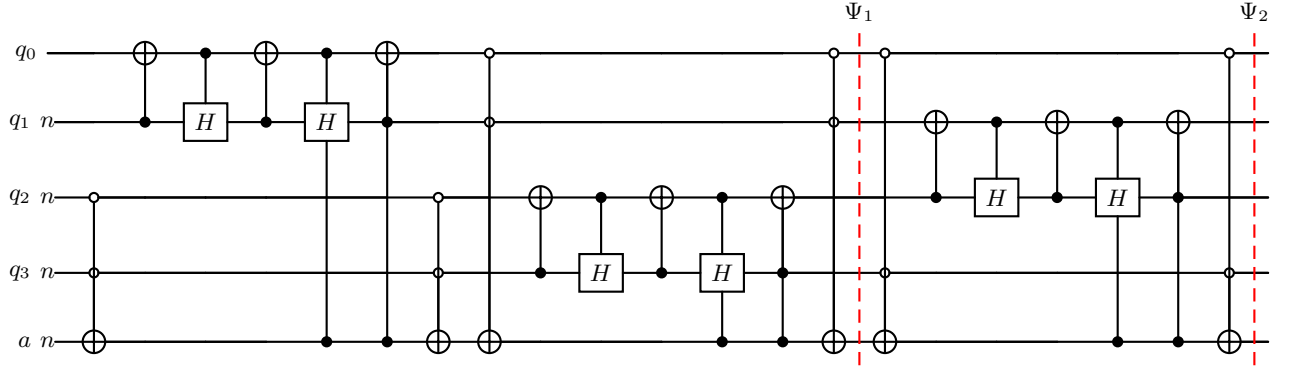
\begin{figure*}
    \centering
    \begin{adjustbox}{max width=\textwidth}
    \begin{quantikz}[transparent, column sep=0.35cm]
        \lstick{$q_0$} & \qw       & \targ{}   & \ctrl{1} & \targ{}   & \ctrl{1}  & \targ{}   & \qw       & \octrl{4} & \qw       & \qw      & \qw       & \qw       & \qw       & \octrl{4} \slice{$\Psi_1$} & \octrl{4} & \qw       & \qw      & \qw       & \qw       & \qw       & \octrl{4} \slice{$\Psi_2$} & \qw \\n        \lstick{$q_1$} & \qw       & \ctrl{-1} & \gate{H} & \ctrl{-1} & \gate{H}  & \ctrl{-1} & \qw       & \octrl{3} & \qw       & \qw      & \qw       & \qw       & \qw       & \octrl{3}                  & \qw       & \targ{}   & \ctrl{1} & \targ{}   & \ctrl{1}  & \targ{}   & \qw                        & \qw \\n        \lstick{$q_2$} & \octrl{2} & \qw       & \qw      & \qw       & \qw       & \qw       & \octrl{2} & \qw       & \targ{}   & \ctrl{1} & \targ{}   & \ctrl{1}  & \targ{}   & \qw                        & \qw       & \ctrl{-1} & \gate{H} & \ctrl{-1} & \gate{H}  & \ctrl{-1} & \qw                        & \qw \\n        \lstick{$q_3$} & \octrl{1} & \qw       & \qw      & \qw       & \qw       & \qw       & \octrl{1} & \qw       & \ctrl{-1} & \gate{H} & \ctrl{-1} & \gate{H}  & \ctrl{-1} & \qw                        & \octrl{1} & \qw       & \qw      & \qw       & \qw       & \qw       & \octrl{1}                  & \qw \\n        \lstick{$a$}   & \targ{}   & \qw       & \qw      & \qw       & \ctrl{-3} & \ctrl{-4} & \targ{}   & \targ{}   & \qw       & \qw      & \qw       & \ctrl{-1} & \ctrl{-2} & \targ{}                    & \targ{}   & \qw       & \qw      & \qw       & \ctrl{-2} & \ctrl{-3} & \targ{}                    & \qw
    \end{quantikz}
    \end{adjustbox}
    \caption{Generalized input-independent quantum circuit corresponding to the four-mode two-photon interferometer. Each optical beam splitter is replaced by the encoded two-photon beam-splitter unit derived in Sec.~IV. Ancilla-assisted spectator-mode identification activates the local transformation only within the corresponding two-photon interference subspace. The circuit reproduces the complete four-mode evolution for all physical input states in the encoded Hilbert space.}
    \label{fig:4mode_circuit}
\end{figure*}

The complete four-mode quantum circuit is constructed by replacing each beam splitter in the interferometer with the two-photon repeating unit derived above. For a beam splitter acting on modes $i$ and $j$, the remaining two mode qubits identify whether the input belongs to the corresponding local two-photon interference subspace. The resulting input-independent circuit, shown in Fig.~\ref{fig:4mode_circuit}, reproduces the four-mode transformation over the complete ten-dimensional two-photon Hilbert space.

Although the generalized circuit shown in Fig.~\ref{fig:4mode_circuit} exactly reproduces the two-photon evolution for arbitrary input states within the encoded Hilbert space, its direct realization using linear optics is experimentally challenging. The presence of several multiqubit-controlled operations leads to a large number of beam splitters, polarizing beam splitters, wave plates, conditional routing elements, and path-relabelling operations. In the present work, we therefore exploit the fact that the input state is fixed in each experimental run and construct a reduced circuit for each of the six inputs,
\begin{equation}
\mathcal{I}_{\mathrm{cf}}
=
\left\{
\ket{1100},
\ket{1010},
\ket{1001},
\ket{0110},
\ket{0101},
\ket{0011}
\right\}.
\label{eq:collision_free_inputs}
\end{equation}

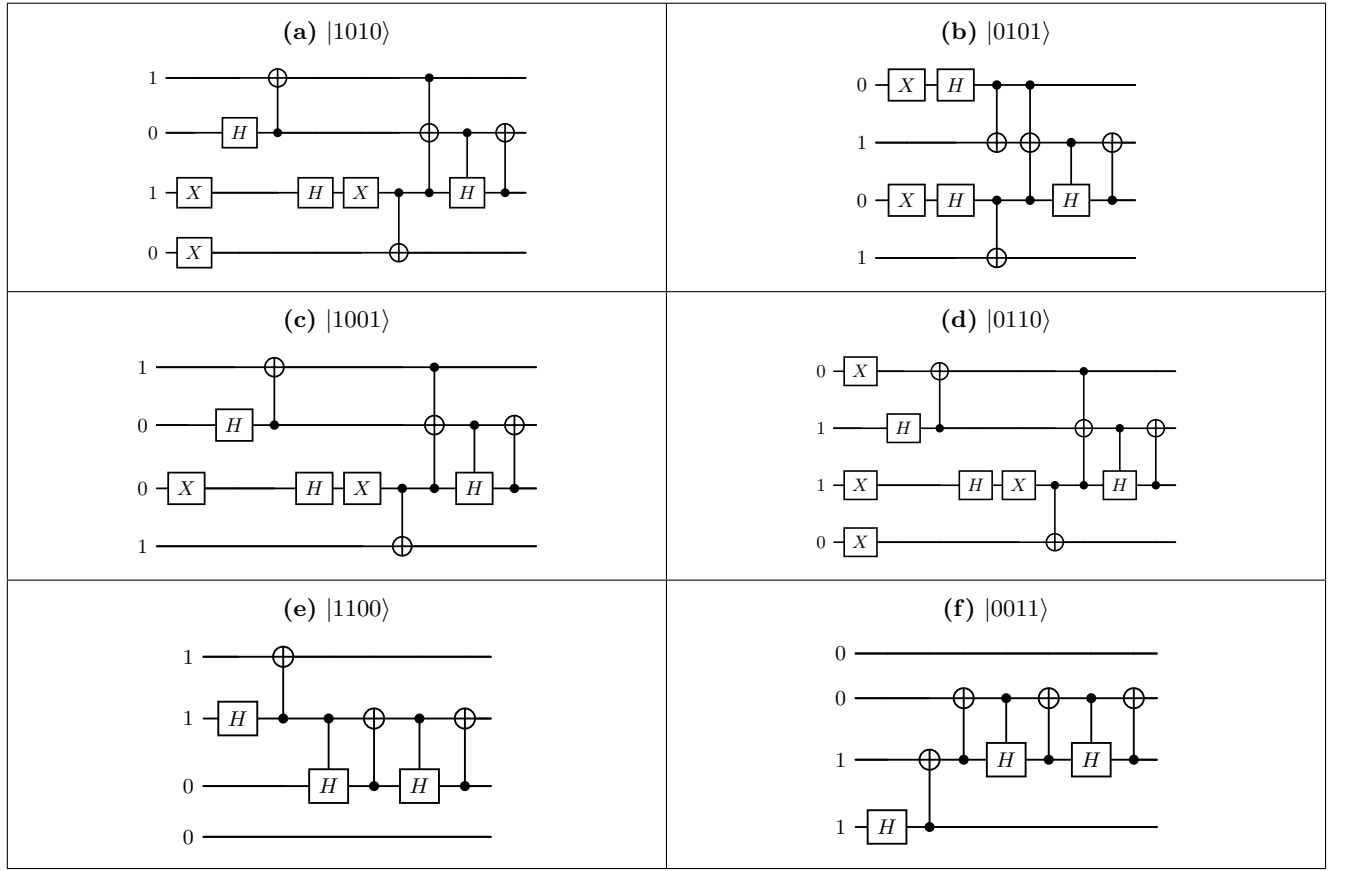
\begin{figure*}[!t]
\centering
\setlength{\tabcolsep}{4pt}
\renewcommand{\arraystretch}{1.1}

\begin{tabular}{|c|c|}
\hline

\begin{minipage}[c][3.8cm][c]{0.47\textwidth}
\centering
\textbf{(a)} $\lvert1010\rangle$\\[2mm]
\begin{adjustbox}{max width=\linewidth,max totalheight=2.8cm}
\begin{quantikz}[column sep=0.18cm]
\lstick{1} & \qw      & \qw      & \targ{}   & \qw      & \qw      & \qw      & \ctrl{1}  & \qw      & \qw       & \qw \\
\lstick{0} & \qw      & \gate{H} & \ctrl{-1} & \qw      & \qw      & \qw      & \targ{}   & \ctrl{1} & \targ{}   & \qw \\
\lstick{1} & \gate{X} & \qw      & \qw       & \gate{H} & \gate{X} & \ctrl{1} & \ctrl{-1} & \gate{H} & \ctrl{-1} & \qw \\
\lstick{0} & \gate{X} & \qw      & \qw       & \qw      & \qw      & \targ{}  & \qw       & \qw      & \qw       & \qw
\end{quantikz}
\end{adjustbox}
\end{minipage}
&
\begin{minipage}[c][3.8cm][c]{0.47\textwidth}
\centering
\textbf{(b)} $\lvert0101\rangle$\\[2mm]
\begin{adjustbox}{max width=\linewidth,max totalheight=2.8cm}
\begin{quantikz}[column sep=0.20cm]
\lstick{0} & \gate{X} & \gate{H} & \ctrl{1} & \ctrl{1}  & \qw      & \qw       & \qw \\
\lstick{1} & \qw      & \qw      & \targ{}  & \targ{}   & \ctrl{1} & \targ{}   & \qw \\
\lstick{0} & \gate{X} & \gate{H} & \ctrl{1} & \ctrl{-1} & \gate{H} & \ctrl{-1} & \qw \\
\lstick{1} & \qw      & \qw      & \targ{}  & \qw       & \qw      & \qw       & \qw
\end{quantikz}
\end{adjustbox}
\end{minipage}
\\
\hline

\begin{minipage}[c][3.8cm][c]{0.47\textwidth}
\centering
\textbf{(c)} $\lvert1001\rangle$\\[2mm]
\begin{adjustbox}{max width=\linewidth,max totalheight=2.8cm}
\begin{quantikz}[column sep=0.18cm]
\lstick{1} & \qw      & \qw      & \targ{}   & \qw      & \qw      & \qw      & \ctrl{1}  & \qw      & \qw       & \qw \\
\lstick{0} & \qw      & \gate{H} & \ctrl{-1} & \qw      & \qw      & \qw      & \targ{}   & \ctrl{1} & \targ{}   & \qw \\
\lstick{0} & \gate{X} & \qw      & \qw       & \gate{H} & \gate{X} & \ctrl{1} & \ctrl{-1} & \gate{H} & \ctrl{-1} & \qw \\
\lstick{1} & \qw      & \qw      & \qw       & \qw      & \qw      & \targ{}  & \qw       & \qw      & \qw       & \qw
\end{quantikz}
\end{adjustbox}
\end{minipage}
&
\begin{minipage}[c][3.8cm][c]{0.47\textwidth}
\centering
\textbf{(d)} $\lvert0110\rangle$\\[2mm]
\begin{adjustbox}{max width=\linewidth,max totalheight=2.8cm}
\begin{quantikz}[column sep=0.18cm]
\lstick{0} & \gate{X} & \qw      & \targ{}   & \qw      & \qw      & \qw      & \ctrl{1}  & \qw      & \qw       & \qw \\
\lstick{1} & \qw      & \gate{H} & \ctrl{-1} & \qw      & \qw      & \qw      & \targ{}   & \ctrl{1} & \targ{}   & \qw \\
\lstick{1} & \gate{X} & \qw      & \qw       & \gate{H} & \gate{X} & \ctrl{1} & \ctrl{-1} & \gate{H} & \ctrl{-1} & \qw \\
\lstick{0} & \gate{X} & \qw      & \qw       & \qw      & \qw      & \targ{}  & \qw       & \qw      & \qw       & \qw
\end{quantikz}
\end{adjustbox}
\end{minipage}
\\
\hline

\begin{minipage}[c][3.8cm][c]{0.47\textwidth}
\centering
\textbf{(e)} $\lvert1100\rangle$\\[2mm]
\begin{adjustbox}{max width=\linewidth,max totalheight=2.8cm}
\begin{quantikz}[column sep=0.22cm]
\lstick{1} & \qw      & \targ{}   & \qw      & \qw       & \qw      & \qw       & \qw \\
\lstick{1} & \gate{H} & \ctrl{-1} & \ctrl{1} & \targ{}   & \ctrl{1} & \targ{}   & \qw \\
\lstick{0} & \qw      & \qw       & \gate{H} & \ctrl{-1} & \gate{H} & \ctrl{-1} & \qw \\
\lstick{0} & \qw      & \qw       & \qw      & \qw       & \qw      & \qw       & \qw
\end{quantikz}
\end{adjustbox}
\end{minipage}
&
\begin{minipage}[c][3.8cm][c]{0.47\textwidth}
\centering
\textbf{(f)} $\lvert0011\rangle$\\[2mm]
\begin{adjustbox}{max width=\linewidth,max totalheight=2.8cm}
\begin{quantikz}[column sep=0.18cm]
\lstick{$0$} & \qw      & \qw      & \qw      & \qw      & \qw      & \qw      & \qw      & \qw \\
\lstick{$0$} & \qw      & \qw      & \targ{}  & \ctrl{1} & \targ{}  & \ctrl{1} & \targ{}  & \qw \\
\lstick{$1$} & \qw      & \targ{}  & \ctrl{-1}& \gate{H} & \ctrl{-1}& \gate{H} & \ctrl{-1}& \qw \\
\lstick{$1$} & \gate{H} & \ctrl{-1}& \qw      & \qw      & \qw      & \qw      & \qw      & \qw
\end{quantikz}
\end{adjustbox}
\end{minipage}
\\
\hline

\end{tabular}
\caption{
Optimized quantum circuits corresponding to the six  input states of the four-mode interferometer. Qubits $q_0$, $q_1$, and $q_3$ represent path qubits, whereas $q_2$ represents the polarization qubit. Here, $H$ and $X$ denote the Hadamard and Pauli-$X$ gates, respectively.
}
\label{fig:optimized_circuits}
\end{figure*}

Since the input state is fixed in each measurement, inactive controlled operations and redundant gate sequences can be removed without changing the corresponding output state. The six resulting input-specific circuits, shown in Fig.~\ref{fig:optimized_circuits}, are considerably simpler and are directly amenable to linear-optical implementation.

\section{Experimental realization}
\label{Expt}
\subsection{Linear-Optical Implementation}

Single-photon implementations of multiqubit logic using polarization and
spatial degrees of freedom have previously been demonstrated
\cite{knill2001} \cite{pavesi, kok2005linear, kagalwala, shen}. Single-photon-based architectures have also been explored for scalable
boson sampling, including implementations based on time-bin encoding
and programmable optical networks \cite{he2017timebin}. In the present
experiment, the four-qubit logical register is encoded using three path qubits
and one polarization qubit of a heralded single photon. The logical qubits
$q_0$, $q_1$, and $q_3$ are represented by distinct spatial modes, whereas
$q_2$ is encoded in the horizontal ($H$) and vertical ($V$) polarization
states of the photon. The optical network is assembled on a compact
fiber-bench platform, providing a stable and compact realization without
free-space steering mirrors.

The optimized quantum circuits shown in Fig. \ref{fig:optimized_circuits} are realized using passive linear-optical
elements, including balanced beam splitters (BSs), polarizing beam splitters
(PBSs), and half-wave plates (HWPs). A Hadamard operation on a path-encoded
qubit is implemented by a balanced beam splitter, which coherently mixes the
corresponding spatial modes. For the polarization qubit, the Hadamard
operation is realized using a HWP oriented at $22.5^\circ$, while a
Pauli-$X$ operation is implemented using a HWP at $45^\circ$, which
interchanges the horizontal and vertical polarization components.

Controlled operations between the path and polarization degrees of freedom are
implemented using PBSs and HWPs. A PBS separates the horizontal and vertical
polarization components into different spatial channels, after which HWPs
placed in the selected paths perform the required path-dependent polarization
transformations. Controlled operations acting solely on path qubits are
implemented through appropriate relabelling of spatial modes. The
correspondence between the quantum-circuit operations and their optical
implementations is summarized in Table~\ref{tab:gate_mapping}.

\begin{table}[H]
\caption{Correspondence between quantum gates and optical elements.}
\label{tab:gate_mapping}
\begin{ruledtabular}
\begin{tabular}{lc}
\hline
Quantum gate & Optical implementation \\
\hline
$H_{\mathrm{path}}$ & 50:50 Beam splitter \\
$H_{\mathrm{pol}}$ & HWP ($22.5^\circ$) \\
$X_{\mathrm{pol}}$ & HWP ($45^\circ$) \\
$\mathrm{CNOT}_{\mathrm{pol}\rightarrow\mathrm{path}}$ & PBS \\
$\mathrm{CNOT}_{\mathrm{path}\rightarrow\mathrm{path}}$ & Relabelling path \\
\hline
\end{tabular}
\end{ruledtabular}
\end{table}

The same compact fiber-bench platform is used to realize all six optimized circuits using two experimental configurations. The first configuration implements the input states \(\{\ket{1010}, \ket{1001}, \ket{0110}, \ket{0101}\}\), while the second implements \(\ket{0011}\) and \(\ket{1100}\). The state \(\ket{1100}\) is realized by injecting the photon through the alternate input port. The required transformation for each input state is obtained by configuring the corresponding arrangement of beam splitters, polarizing beam splitters, and half-wave plates according to the optimized circuit shown in Fig.~\ref{fig:optimized_circuits}. The complete experimental realization of the optical network and the heralded single-photon source are described in the following subsections.

\subsection{Heralded single-photon source}
The heralded single-photon source is generated through collinear type-II
quasi-phase-matched spontaneous parametric down-conversion in a PPKTP crystal
placed inside a polarization Sagnac interferometer
\cite{Sagnac1,Sagnac2,Sagnac3}.
A continuous-wave 405~nm laser (Toptica TopMode) pumps a 30-mm-long Type-II PPKTP crystal (Raicol, poling period $10~\mu$m) arranged in a Sagnac interferometer, generating degenerate signal and idler photon pairs at 810~nm. Degenerate phase matching is achieved by temperature tuning of the crystal to approximately $24.5^\circ$C. Under the operating conditions used in the experiment, approximately $3.2\times10^{6}$ counts/s are recorded in both the signal and idler arms, with a coincidence rate of about $7.5\times10^{5}$ counts/s within a coincidence window of 2~ns, corresponding to a heralding efficiency of approximately $25\%$. The measured heralded second-order correlation at zero delay is $g_h^{(2)}(0)\approx0.032$, confirming the high single-photon purity of the source. The heralding and output photons are detected using silicon avalanche photodiode single-photon counting modules (Excelitas SPCM-AQRH-WX-FC), with photon arrival times recorded using a Swabian Instruments Time Tagger Ultra. 
\begin{figure}[t]
    \centering
    \includegraphics[width=\columnwidth]{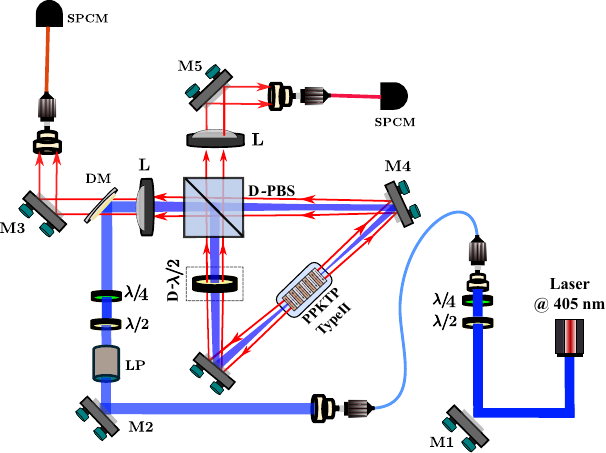}
\caption{Schematic of the experimental setup for heralded single-photon generation. The remaining optical arrangement constitutes a standard type-II Sagnac interferometer. $\lambda/2$, half-wave plates; $\lambda/4$, quarter-wave plates; LP, linear polarizer; L, lens of focal length $20\,\mathrm{cm}$; M1--M5, mirrors; D-PBS, dual-wavelength polarizing beam splitter; DM, dichroic mirror; PPKTP, periodically poled potassium titanyl phosphate crystal; FC, fibre couplers; SPCM, single-photon counting modules. The blue thick line indicates the $405\,\mathrm{nm}$ pump laser, while the red lines indicate the down-converted photons at $810\,\mathrm{nm}$.}
\label{fig:spdc_setup}
    \label{fig:spdc_setup}
\end{figure}

\subsection{Optical realization of the optimized circuits}
\begin{figure}[!t]
    \centering
    \includegraphics[
        width=\columnwidth,
        trim={1.5cm 3.5cm 1.5cm 4.5cm},
        clip
    ]{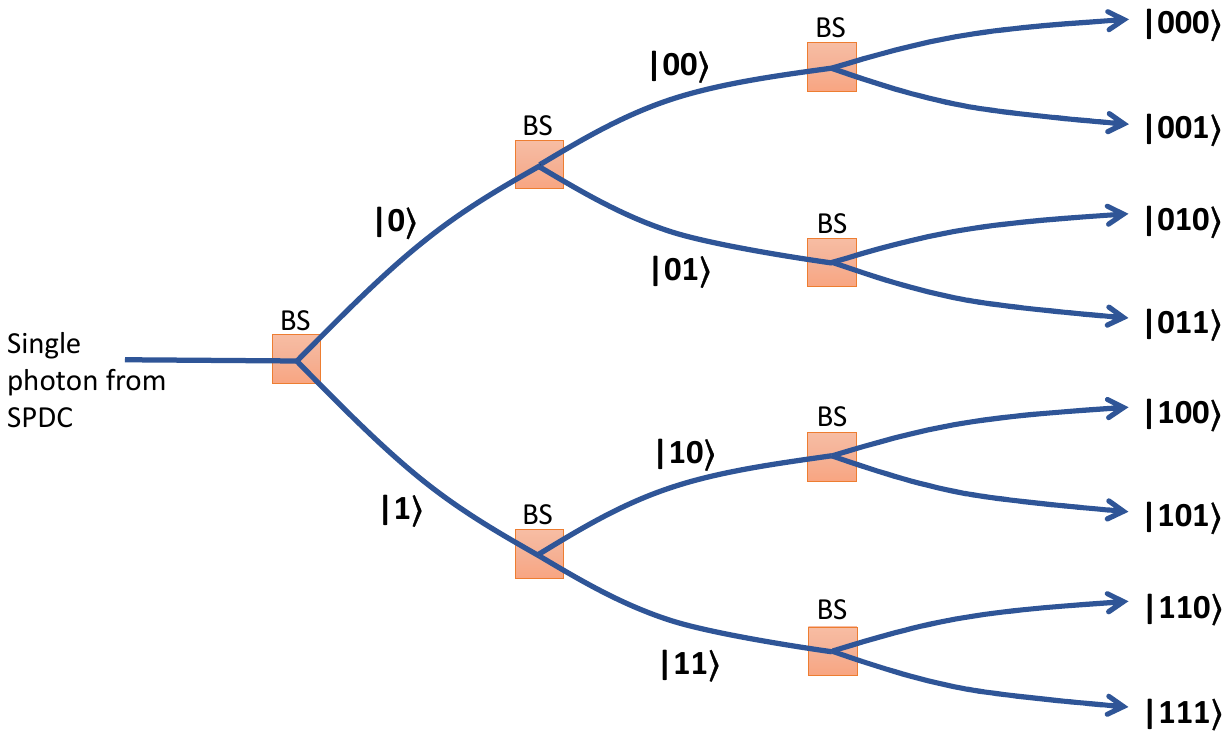}
    \caption{Three-layered balanced beam-splitter network mapping a single photon in eight spatial modes onto the computational basis states $\ket{000}$--$\ket{111}$ of three path-encoded qubits. Each binary label
    records the sequence of path choices through the network.}
    \label{fig:eightfock_encoding}
\end{figure}

The two-qubit Fock-to-path encoding demonstrated using three balanced beam splitters in Ref.\cite{shafi} is extended here to three path qubits. Figure~\ref{fig:eightfock_encoding} illustrates the corresponding binary-tree encoding, in which the two spatial directions at each layer of Binary tree, a qubit is assigned. The eight output rails are thereby labelled by the
computational basis states $\ket{000}$--$\ket{111}$. For the input-specific experimental configurations, the heralded signal photon is coupled directly into the corresponding rail of the path-encoded register. Polarization encodes the fourth qubit ($q_2$), with $\ket{0}_{p}\equiv\ket{H}$ and $\ket{1}_{p}\equiv\ket{V}$.
   
 \begin{figure}[!t]
    \centering
    \begin{subfigure}{\linewidth}
        \centering
        \includegraphics[width=1\linewidth]
        {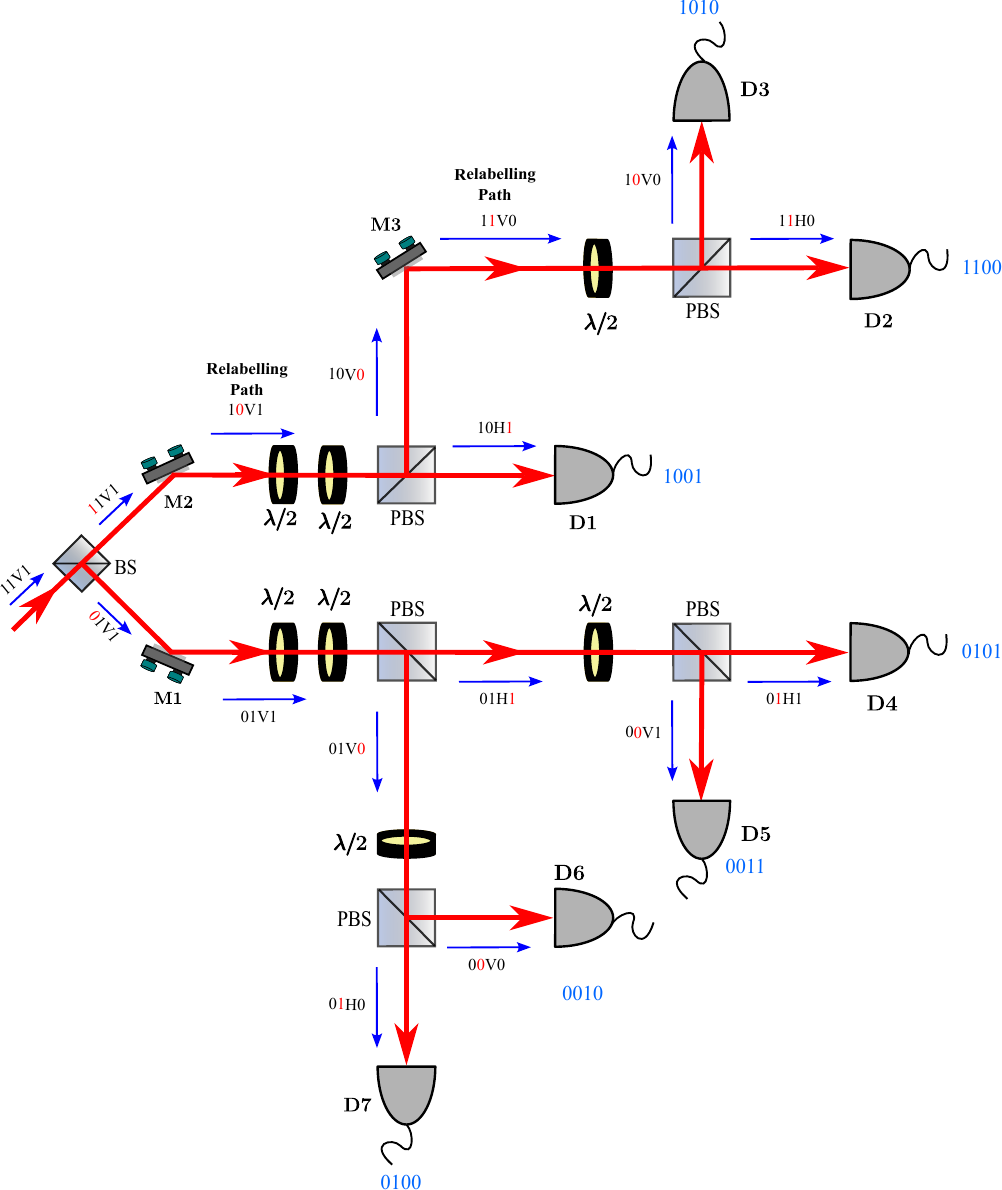}
        \caption{}
        \label{fig:optical_bench1}
    \end{subfigure}
    
    \vspace{1cm}
    
    \begin{subfigure}{\linewidth}
        \centering
        \includegraphics[width=1\linewidth]
        {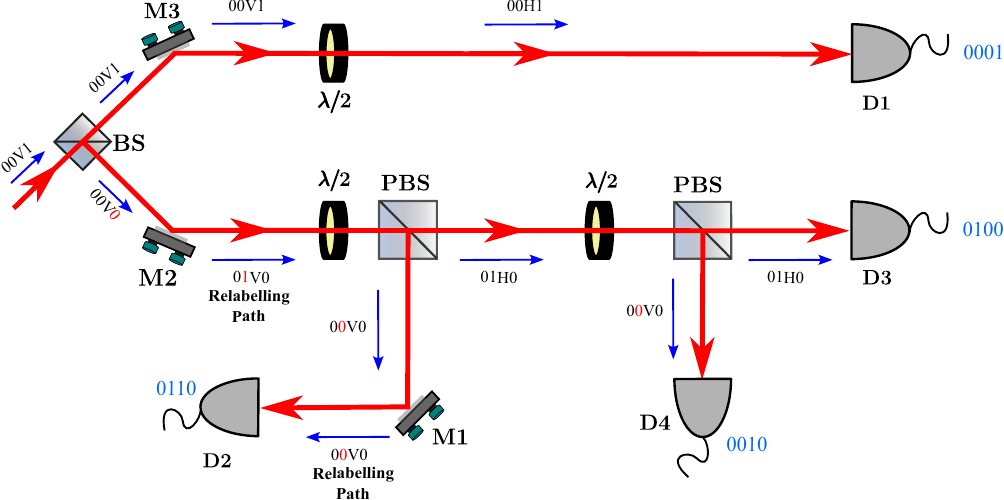}
        \caption{}
        \label{fig:optical_bench2}
    \end{subfigure}
    \caption{Schematic of the experimental configurations used to prepare and
    measure (a) the input states $\ket{1010}$, $\ket{1001}$, $\ket{0110}$,
    and $\ket{0101}$, and (b) the input states $\ket{0011}$ and
    $\ket{1100}$. The red lines indicate the propagation paths of the
    heralded signal photon. The labels 1, 2, and 4 identify the spatial
    paths, while \(V\) denotes vertical polarization. Half-wave plates
    (\(\lambda/2\)), beam splitters (BSs), and polarizing beam splitters
    (PBSs) are represented using standard notation. D1--D7 denote the
    single-photon counting modules (SPCMs).}
    \label{fig:optical_bench_setup}
\end{figure}
Figure~\ref{fig:optical_bench1} shows the experimental setup used to implement the input states \(1010\), \(0101\), \(1001\), and \(0110\). The labeling shown corresponds to the input state \(0101\), and the corresponding quantum circuit is shown in Fig.\ref{fig:optimized_circuits} (b). We begin from the state obtained after applying the initial NOT gates in the circuit. For the input state \(0101\), the application of these NOT gates transforms the state to \(\ket{1111}\), thereby selecting the encoded path state \(\ket{111}\) as the input to the optical network. Experimentally, this is realized by injecting a vertically polarized photon into the interferometer through the input port corresponding to the state \(\ket{11V1}\). The second HWPs following mirrors M1 and M2 are set to \(45^\circ\) for the input states \(1010\), \(1001\), and \(0110\), and to \(0^\circ\) for the input state \(0101\). All the remaining HWPs are maintained at \(22.5^\circ\), thereby implementing Hadamard operations on the polarization qubit.

Figure~\ref{fig:optical_bench2} shows the experimental setup used to implement the input states \(1100\) and \(0011\). The labeling shown corresponds to the input state \(0011\), where the encoded path state is \(\ket{001}\) and the polarization qubit is represented by a vertically polarized photon. In this configuration, the HWP following mirror M3 is set to \(45^\circ\) for the input state \(0011\) and to \(0^\circ\) for the input state \(1100\), while all the remaining HWPs are maintained at \(22.5^\circ\). For the input state \(1100\), the photon is injected through the other input port.
Each output is coupled into a multimode fiber (M42L01) using a fiber-port collimator (PAF2-B), with a coupling efficiency exceeding \(90\%\). The output channels are detected using single-photon counting modules (SPCMs), labelled
D1--D7 in Fig.~\ref{fig:optical_bench1} and D1--D4 in Fig.~\ref{fig:optical_bench2}. The photon arrival times are recorded using a Time Tagger, which performs the coincidence measurements.

\begin{figure*}[t]
    \centering
    \includegraphics[width=0.95\textwidth]{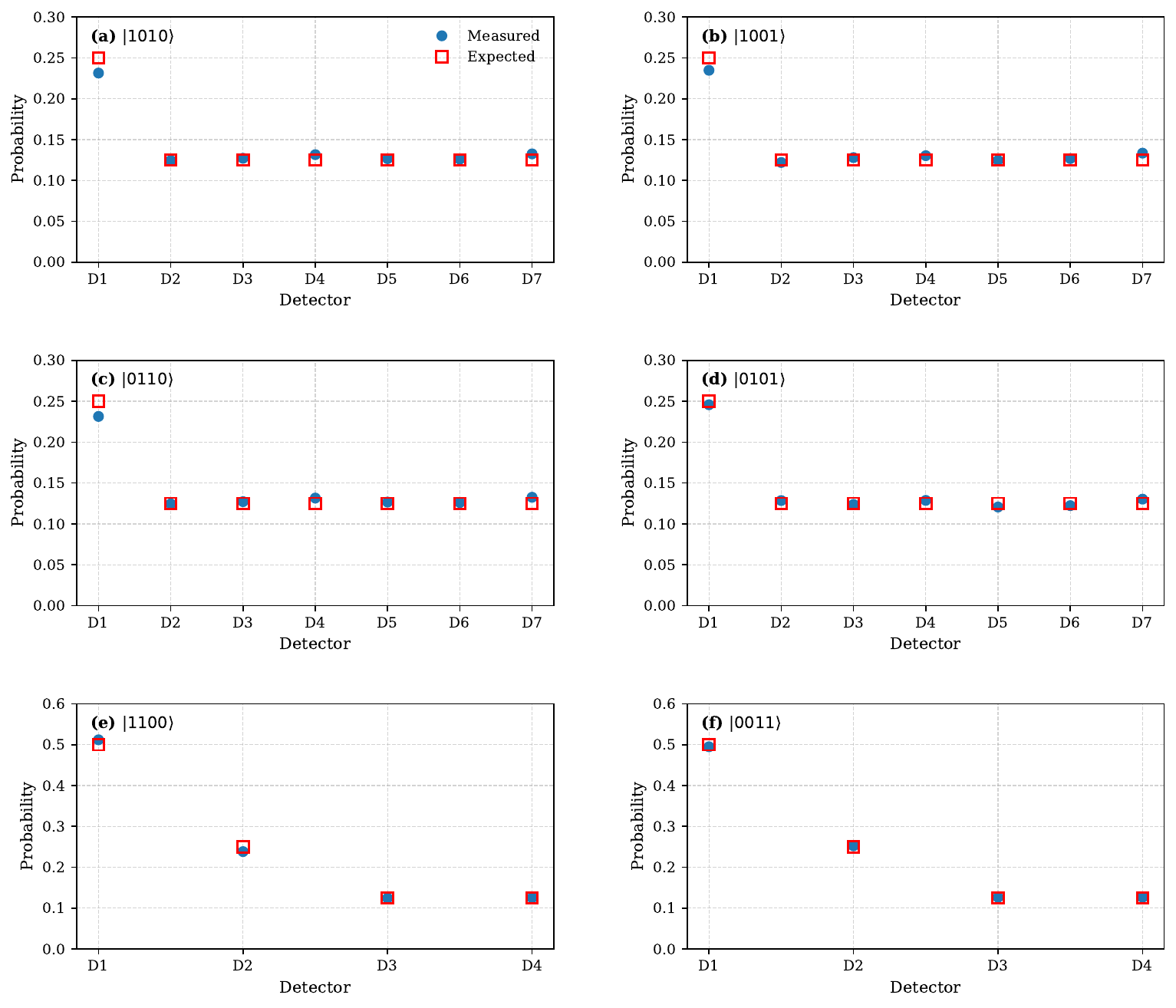}
    \caption{Experimental and theoretical output probability distributions for
the six input states: (a) $\ket{1010}$, (b) $\ket{1001}$,
(c) $\ket{0110}$, (d) $\ket{0101}$, (e) $\ket{1100}$, and
(f) $\ket{0011}$. Blue circles denote the experimentally measured
probabilities, and red squares denote the theoretical predictions.
D1--D7 denote the single-photon counting modules (SPCMs) used in
panels (a)--(d), while D1--D4 denote those used in panels (e) and (f).}
    \label{fig:prob_plots}
\end{figure*}

\subsection{Measurement procedure}
For each input state, coincidence events between the idler herald detector and
the signal-photon output detectors were recorded using a coincidence window of
$2\,\mathrm{ns}$. The measurement was repeated 15 times, with an acquisition
time of 25~s for each repetition. The output channels were recorded sequentially under identical experimental conditions. The
measurements employed seven output channels (D1--D7) for the input states
$\ket{1010}$, $\ket{1001}$, $\ket{0110}$, and $\ket{0101}$, and four output
channels (D1--D4) for $\ket{1100}$ and $\ket{0011}$.

The experimental probability associated with output channel $j$ was obtained
by normalizing the corresponding coincidence count as

\begin{equation}
P_j^{\mathrm{exp}}
=
\frac{C_j}{\displaystyle\sum_k C_k},
\label{eq:experimental_probability}
\end{equation}
where $C_j$ denotes the coincidence count recorded in output channel $j$, and
the summation extends over all monitored output channels for the corresponding
input state. Thus, $P_j^{\mathrm{exp}}$ represents the conditional probability
of detecting the heralded signal photon in output channel $j$, with
\begin{equation}
\sum_j P_j^{\mathrm{exp}}=1.
\end{equation}

The reported probability distributions were obtained by averaging the 15
repeated measurements. Statistical uncertainties in the squared classical
fidelity and total variation distance were estimated by bootstrap resampling
of these measurements. The reconstructed experimental probability
distributions are compared with the corresponding theoretical predictions in
the following subsection.

\subsection{Results and discussion}
\label{sec:results}

The measurements were performed to verify whether the single-photon
path--polarization encoding reproduces the output statistics expected from
two-photon interference in a four-mode bosonic interferometer. For each of the
six  input configurations, the corresponding optimized
linear-optical circuit was implemented and the output probability
distribution was reconstructed from heralded coincidence measurements. The
experimentally obtained distributions were then compared with the theoretical
boson-sampling probabilities in order to assess the accuracy of the encoding
scheme and the overall optical implementation.

Figure~\ref{fig:prob_plots} presents the experimentally measured and
theoretically predicted output probability distributions.
reconstructed and theoretically predicted output probability distributions aforementioned six input  configurations of the four-mode interferometer.
For each input state, the measured probability distribution closely follows
the expected theoretical trend, with the dominant output events occurring at
the predicted output configurations. The overall agreement between experiment
and theory demonstrates that the optimized path--polarization encoding
faithfully reproduces the output statistics of the corresponding two-photon
bosonic interferometer.

To quantify the agreement between the experimentally reconstructed and
theoretically predicted probability distributions, we evaluate both the
classical fidelity and the total variation distance. The classical fidelity
measures the overlap between the two distributions, whereas the total
variation distance directly quantifies their absolute discrepancy. These
quantities are defined as
\begin{equation}
F
=
\left(
\sum_i
\sqrt{
P_i^{\mathrm{exp}}P_i^{\mathrm{th}}
}
\right)^2.
\label{eq:classical_fidelity}
\end{equation}
and
\begin{equation}
D_{\mathrm{TV}}
=
\frac{1}{2}
\sum_i
\left|
P_i^{\mathrm{exp}}-P_i^{\mathrm{th}}
\right|.
\label{eq:tvd}
\end{equation}
Here, $F=1$ and $D_{\mathrm{TV}}=0$ correspond to perfect agreement between
the experimental and theoretical probability distributions.
\begin{table}[t]
\caption{Squared classical fidelity and total variation distance between the
experimentally reconstructed and theoretically predicted output probability
distributions for the six  input states. The quoted
uncertainties are one-standard-error bootstrap estimates obtained from
15 repeated acquisitions.}
\label{tab:distribution_metrics}
\centering
\begin{ruledtabular}
\begin{tabular}{ccc}
Input state & Fidelity ($F$) & TVD ($D_{\mathrm{TV}}$) \\
\hline
$\ket{1010}$ &
$0.999429 \pm 1.5\times10^{-5}$ &
$0.01905 \pm 2.7\times10^{-4}$ \\

$\ket{1001}$ &
$0.999526 \pm 1.3\times10^{-5}$ &
$0.01856 \pm 2.0\times10^{-4}$ \\

$\ket{0110}$ &
$0.999429 \pm 1.5\times10^{-5}$ &
$0.01905 \pm 2.7\times10^{-4}$ \\

$\ket{0101}$ &
$0.999819 \pm 6.0\times10^{-6}$ &
$0.01254 \pm 2.1\times10^{-4}$ \\

$\ket{1100}$ &
$0.999791 \pm 1.0\times10^{-5}$ &
$0.01216 \pm 3.9\times10^{-4}$ \\

$\ket{0011}$ &
$0.999973 \pm 4.0\times10^{-6}$ &
$0.00517 \pm 3.7\times10^{-4}$ \\
\end{tabular}
\end{ruledtabular}
\end{table}
As summarized in Table~\ref{tab:distribution_metrics}, all six input configurations exhibit squared classical fidelities above $0.9994$. The corresponding total variation distances lie between approximately $0.005$ and $0.019$, indicating that the total probability-weight discrepancy between experiment and theory remains below $2\%$ for all
investigated input states.

\section{Conclusion}
\label{Conc}

We have presented a gate-based quantum-circuit framework for boson sampling by mapping a bosonic Hilbert spaces to the  multi-qubit state. By mapping bosonic beam-splitter transformations onto gate operation on the qubit states, we constructed explicit quantum circuits for one- and two-photon interference processes using standard quantum gates. Furthermore, we developed a repeating-unit architecture that enables the systematic construction of two-photon $n$-mode boson sampling circuits, providing a scalable approach for larger interferometric networks.

To demonstrate the practicality of the framework, we constructed the quantum-circuit representation of a four-mode two-photon boson-sampling interferometer and derived optimized input-specific circuits suitable for linear-optical implementation. These circuits were experimentally realized using a heralded single photon encoded in its path and polarization degrees of freedom to generate four-qubit state, allowing the emulation of the corresponding two-photon interference dynamics within a compact multi-degree-of-freedom photonic platform.

The experimentally reconstructed output probability distributions for all six input states showed excellent agreement with the theoretical boson-sampling predictions. The measured squared classical fidelities exceeded 0.9994 for every input configuration, while the total variation distances remained below 0.02, confirming that the proposed encoding accurately reproduces the expected bosonic interference statistics. The small residual discrepancies are consistent with experimental imperfections such as finite optical alignment accuracy, limited component extinction ratios, and statistical fluctuations in the coincidence measurements. More generally, the influence of experimental imperfections on the scalability and computational behavior of boson-sampling systems has been extensively investigated \cite{rohde2012error}. Photon loss and partial distinguishability also constitute important limitations for large-scale boson sampling, since sufficiently strong
imperfections can reduce the complexity of classically simulating the resulting distributions
\cite{renema2018efficient,garciapatron2019simulating}. The presented framework establishes a direct connection between bosonic linear-optical networks and quantum-circuit representations, enabling boson sampling to be studied within the standard quantum computing paradigm. Beyond photonic implementations, the circuit-based formulation provides a platform-independent description that can, in principle, be realized on other qubit-based quantum hardware, including superconducting circuits, trapped ions, and neutral atoms.

\begin{acknowledgments}
We thank Mr Kunal Shukla, Mr Kanad Sengupta and Mr Anirudh Verma for insightful discussion on the theoretical aspect of the work. We acknowledge funding support from the National Quantum Mission, an initiative of the Department of Science and Technology, Govt. of India.
\end{acknowledgments}

\section{Appendix} 
\subsection{Proof of the Uniqueness of the Proposed Encoding}

\noindent\textbf{Proposition}
\vspace{0.3cm}

Consider the two-photon Fock basis over $n$ optical modes,
\[
\mathcal{F}_{2,n}
=
\left\{
(m_1,\ldots,m_n):
m_i\in\{0,1,2\},
\sum_{i=1}^{n}m_i=2
\right\}.
\]

We restrict attention to encodings satisfying the following conditions:
(i) each optical mode is represented by one qubit;
(ii) the same mode-independent repeating unit (Fig.~\ref{fig:repeating_unit}) is used to implement the
beam-splitter transformation between every pair of modes. Under these assumptions, the encoding is
unique up to bitwise complementation of all mode qubits and is given by
\[
x_i=
\begin{cases}
1,&m_i>0,\\
0,&m_i=0.
\end{cases}
\]

\noindent\textbf{Proof}
\vspace{0.3cm}

Consider the Fock state
\[
|2,0,\ldots,0\rangle,
\]
and let its encoding be
\[
x_1x_2\cdots x_n,\qquad x_i\in\{0,1\}.
\]

Since the same repeating unit simulates a beam splitter acting on any pair of modes, the logical representation of the local occupation $(2,0)$ must be identical irrespective of the second mode chosen. Thus,
\[
(x_1,x_2)
=
(x_1,x_3)
=
\cdots
=
(x_1,x_n),
\]
which immediately implies
\[
x_2=x_3=\cdots=x_n.
\]

Since the occupied mode must be distinguishable from the unoccupied modes,
\[
x_1\neq x_2.
\]
Therefore, only two encodings are possible:
\[
100\cdots0,
\qquad
011\cdots1,
\]
which differ only by a global bit complement. Without loss of generality, we choose
\[
|2,0,\ldots,0\rangle
\longrightarrow
100\cdots0.
\]
By symmetry, if two photons occupy the $i$-th mode,
\[
|0,\ldots,2,\ldots,0\rangle
\]
is encoded by placing a logical $1$ at the $i$-th position and logical $0$ elsewhere.

Next, consider the state
\[
|1,0,1,0,\ldots,0\rangle,
\]
whose encoding is
\[
x_1x_2x_3x_4\cdots x_n.
\]
The local occupation $(1,0)$ appears whenever a beam splitter acts on the occupied first mode and any unoccupied mode. Since the same repeating unit is used for every beam splitter,
\[
(x_1,x_2)
=
(x_1,x_4)
=
\cdots,
\]
which gives
\[
x_2=x_4=\cdots.
\]
Similarly, considering beam splitters acting on the occupied third mode,
\[
(x_3,x_2)
=
(x_3,x_4)
=
\cdots,
\]
implies
\[
x_2=x_4=\cdots.
\]
Hence all unoccupied modes must have the same logical value. Since the occupied modes must be distinguishable from the unoccupied modes,
\[
x_1=x_3\neq x_2.
\]
Therefore, the only possible encodings are
\[
1010\cdots0,
\qquad
0101\cdots1,
\]
which again differ only by a global bit complement. Choosing the former gives
\[
|1,0,1,0,\ldots,0\rangle
\longrightarrow
1010\cdots0.
\]
By symmetry, every occupied mode is assigned the logical value $1$ and every unoccupied mode the logical value $0$. Therefore,
\[
x_i=
\begin{cases}
1,&m_i>0,\\
0,&m_i=0,
\end{cases}
\]
which proves the proposed encoding.


\subsection{Three-photon $n$ mode generalized scheme}

In this appendix, we extend the proposed construction to boson sampling with three indistinguishable photons distributed across $n$ spatial modes. 

\subsubsection{Three-Photon Beam-Splitter Transformation}

We first consider the action of a balanced beam splitter on three photons distributed across two spatial modes. For a fixed total photon number $N=3$, the allowed two-mode Fock states are
\begin{equation}
\left\{
\ket{3,0},\,
\ket{0,3},\,
\ket{2,1},\,
\ket{1,2}
\right\}.
\end{equation}
The corresponding Hilbert space is therefore four-dimensional and can be represented using two qubits. We choose the encoding as shown in Table \ref{tab:threephoton},
\begin{table}[h]
\centering
\caption{Encoding for the three-photon sector.}
\label{tab:threephoton}
\begin{tabular}{cc}
\hline
Fock State & Computational Basis State \\
\hline
$\ket{3,0}$ & $\ket{00}$ \\
$\ket{0,3}$ & $\ket{10}$ \\
$\ket{2,1}$ & $\ket{11}$ \\
$\ket{1,2}$ & $\ket{01}$ \\
\hline
\end{tabular}
\end{table}
where the ordering is chosen to facilitate the circuit representation of the beam-splitter transformation.

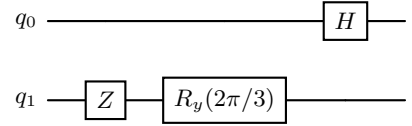
\begin{figure}[htpb]
    \centering
    \begin{quantikz}
        \lstick{$q_0$} & & & \gate{H} & \\
        \lstick{$q_1$} & \gate{Z} & \gate{R_y(2\pi/3)} & & 
    \end{quantikz}
    \caption{Quantum circuit implementing the encoded three-photon beam splitter transformation.}
    \label{fig:three_photon_circuit}
\end{figure}

Collecting the coefficients in the same ordered basis, the beam-splitter transformation in the three-photon sector is
\begin{equation}
U^{(3)}
=
\frac{1}{2\sqrt{2}}
\begin{pmatrix}
1 & 1 & \sqrt{3} & \sqrt{3}\\
1 & -1 & -\sqrt{3} & \sqrt{3}\\
\sqrt{3} & -\sqrt{3} & 1 & -1\\
\sqrt{3} & \sqrt{3} & -1 & -1
\end{pmatrix}.
\label{eq:three_photon_unitary}
\end{equation}
The corresponding quantum circuit implementing this unitary is shown in Fig.~\ref{fig:three_photon_circuit}.

\subsubsection{One-, Two- and Three-Photon Inputs}


To extend the construction to a multimode interferometer, it is necessary to consider the complete local Hilbert space encountered by an individual beam splitter. Although the total photon number of the system is fixed at three, the two modes interacting at a given beam splitter may contain zero, one, two, or all three photons, depending on the distribution of the remaining photons among the other modes. The local two-mode Hilbert space must therefore include all Fock states with total occupation number up to three.

The corresponding fock basis is
\begin{equation}
\mathcal{B}_{\leq 3}
=
\left\{
\begin{aligned}
&\ket{3,0},\,
\ket{0,3},\,
\ket{2,1},\,
\ket{1,2},\,
\ket{2,0},\\
&\ket{0,2},\,
\ket{1,1},\,
\ket{1,0},\,
\ket{0,1},\,
\ket{0,0}
\end{aligned}
\right\}.
\label{eq:three_photon_truncated_basis}
\end{equation}
The dimension of this truncated Hilbert space is ten. Since
\begin{equation}
2^3=8<10\leq16=2^4,
\end{equation}
a minimum of four qubits is required to embed this Hilbert space.

\subsubsection*{Encoding Scheme}

\noindent The encoding is divided into two parts:
\vspace{0.3cm}

\noindent A. The middle two qubits encode the spatial occupation of the two modes as shown in Table \ref{tab:threephoton}

\vspace{0.2cm}
\noindent B. The first and last qubits act as a photon-number register that distinguishes the excitation sectors according to
\[
\begin{aligned}
00 &\rightarrow \text{vacuum and single-photon sector}, \\
01 &\rightarrow \text{two-photon sector}, \\
10 &\rightarrow \text{three-photon sector}.
\end{aligned}
\]
Their purpose is to distinguish between different total excitation subspaces so that controlled operations can be applied selectively in the circuit. Thus each Fock state $|n_1, n_2\rangle$ is mapped according to
\[
|n_1, n_2\rangle \longrightarrow |r_1 q_1 q_2\, r_2\rangle,
\]
where $q_1 q_2$ encode the spatial distribution and $r_1 r_2$ specify the total
photon-number sector. The complete encoding is given in Table \ref{tab:full_encoding}
\begin{table}[h]
\centering
\caption{Encoding for the Hilbert space with a maximum photon number of three.}
\label{tab:full_encoding}
\begin{tabular}{cc}
\hline
Fock State & Computational Basis State \\
\hline
$\ket{3,0}$ & $\ket{1000}$ \\
$\ket{0,3}$ & $\ket{1100}$ \\
$\ket{2,1}$ & $\ket{1110}$ \\
$\ket{1,2}$ & $\ket{1010}$ \\
$\ket{2,0}$ & $\ket{0101}$ \\
$\ket{0,2}$ & $\ket{0011}$ \\
$\ket{1,1}$ & $\ket{0111}$ \\
$\ket{1,0}$ & $\ket{0100}$ \\
$\ket{0,1}$ & $\ket{0010}$ \\
$\ket{0,0}$ & $\ket{0000}$ \\
\hline
\end{tabular}
\end{table}

\begin{figure}[htpb]
    \centering
    \resizebox{0.5\textwidth}{!}{%
        \begin{quantikz}
            \lstick{$q_0$} & \ctrl{2}  & \ctrl{2}           & \ctrl{1}  & \octrl{1} & \octrl{2} & \octrl{1} &           &           & \\
            \lstick{$q_1$} &           &                    & \gate{H}  & \targ{}   & \ctrl{1}  & \targ{}   & \ctrl{1}  & \targ{}   & \\
            \lstick{$q_2$} & \gate{Z}  & \gate{R_y(2\pi/3)} &           & \ctrl{-1} & \gate{H}  & \ctrl{-1} & \gate{H}  & \ctrl{-1} & \\
            \lstick{$q_3$} &           &                    &           &           &           &           & \ctrl{-1} & \ctrl{-2} &
        \end{quantikz}%
    }
    \caption{Quantum circuit implementing the beam splitter transformation in the Hilbert space with maximum photon number equal to three.}
    \label{fig:max_three_photon_circuit}
\end{figure}
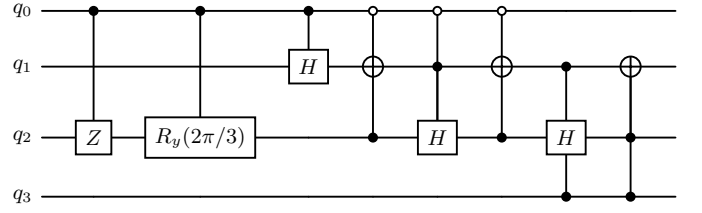

All remaining computational basis states lie outside this Hilbert space and are therefore discarded as unphysical. The corresponding quantum circuit implementing this encoded transformation is shown in Fig.~\ref{fig:max_three_photon_circuit}.

\subsubsection*{Genralized Scheme}

We now extend the three-photon construction to an arbitrary (n)-mode linear-optical interferometer. For a fixed total photon number (N=3), the occupation numbers satisfy
\begin{equation}
\sum_{i=1}^{n} n_i = 3.
\label{eq:three_constraint}
\end{equation}
The allowed Fock configurations belong to one of the following three classes:
\begin{enumerate}
    \item All three photons occupy the same mode,
    \begin{equation}
    |0,\ldots,3_k,\ldots,0\rangle,\qquad k=1,\ldots,n,
    \end{equation}
    \item Two photons occupy one mode and one photon occupies another,
    \begin{equation}
    |0,\ldots,2_i,\ldots,1_j,\ldots,0\rangle,\qquad i\neq j,
    \end{equation}
    \item Each photon occupies a different mode,
    \begin{equation}
    |0,\ldots,1_i,\ldots,1_j,\ldots,1_k,\ldots,0\rangle,\qquad i\neq j\neq k.
    \end{equation}
\end{enumerate}

To obtain an encoding that extends directly with the number of spatial modes, we assign a two-qubit register to each mode. Since each mode can contain $0$, $1$, $2$, or $3$ photons,
\begin{equation}
n_i\in\{0,1,2,3\},
\end{equation}
its four possible occupation numbers are represented using the Gray-code encoding~\cite{sawaya2020}
\begin{table}[h]
\centering
\caption{Gray-code encoding of the photon occupation number of a spatial mode.}
\label{tab:scalable_encoding}
\begin{tabular}{cc}
\hline
Photon Number $n_i$ & Computational Basis State \\
\hline
$0$ & $\ket{00}$ \\
$1$ & $\ket{10}$ \\
$2$ & $\ket{11}$ \\
$3$ & $\ket{01}$ \\
\hline
\end{tabular}
\end{table}

This encoding scheme is designed to ensure scalability. Thus, the occupation of the $i^{\mathrm{th}}$ mode is encoded as
\begin{equation}
n_i \rightarrow |q_{2i-1}q_{2i}\rangle,
\end{equation}
and an $n$-mode interferometer requires
\begin{equation}
2n
\end{equation}
qubits. The corresponding encoded computational basis state is
\begin{equation}
|n_1,n_2,\ldots,n_n\rangle
\longrightarrow
|q_1q_2q_3q_4\cdots q_{2n-1}q_{2n}\rangle.
\end{equation}

\begin{figure*}[t]
\centering
\resizebox{1.00\textwidth}{!}{%
\begin{quantikz}[
    row sep=0.45cm,
    column sep=0.30cm
]
\qw
& \qw
& \octrl{3}
& \targ{}
& \targ{}
& \ctrl{2}
& \ctrl{3}
& \ctrl{3}
& \targ{}
& \ctrl{2}
& \ctrl{2}
& \ctrl{1}
& \octrl{2}
& \octrl{2}
& \octrl{2}
& \qw
& \qw
& \targ{}
& \ctrl{3}
& \ctrl{3}
& \ctrl{2}
& \targ{}
& \targ{}
& \octrl{3}
& \qw
& \qw
\\
\qw
& \targ{}\vqw{2}
& \ctrl{2}
& \ctrl{-1}
& \ctrl{-1}
& \targ{}
& \ctrl{2}
& \targ{}
& \ctrl{-1}
& \qw
& \qw
& \gate{H}
& \targ{}
& \ctrl{1}
& \targ{}
& \ctrl{2}
& \targ{}
& \ctrl{-1}
& \targ{}
& \ctrl{2}
& \targ{}
& \ctrl{-1}
& \ctrl{-1}
& \ctrl{2}
& \targ{}\vqw{2}
& \qw
\\
\qw
& \ocontrol{}
& \qw
& \qw
& \ctrl{-2}
& \ctrl{-1}
& \qw
& \qw
& \qw
& \gate{Z}
& \gate{R_y(2\pi/3)}
& \qw
& \ctrl{-1}
& \gate{H}
& \ctrl{-1}
& \gate{H}
& \ctrl{-1}
& \qw
& \qw
& \qw
& \ctrl{-1}
& \ctrl{-2}
& \qw
& \qw
& \ocontrol{}
& \qw
\\
\qw
& \control{}
& \targ{}
& \ctrl{-3}
& \qw
& \qw
& \targ{}
& \octrl{-2}
& \qw
& \qw
& \qw
& \qw
& \qw
& \qw
& \qw
& \ctrl{-1}
& \ctrl{-2}
& \qw
& \octrl{-2}
& \targ{}
& \qw
& \qw
& \ctrl{-3}
& \targ{}
& \control{}
& \qw
\end{quantikz}%
}
\caption{Repeating quantum circuit unit corresponding to a single beam splitter in the generalized three-photon, \(n\)-mode boson sampling circuit. This block is applied sequentially for each beam splitter in the interferometer decomposition to construct the complete quantum circuit.}
\label{fig:four_qubit_circuits}
\end{figure*}
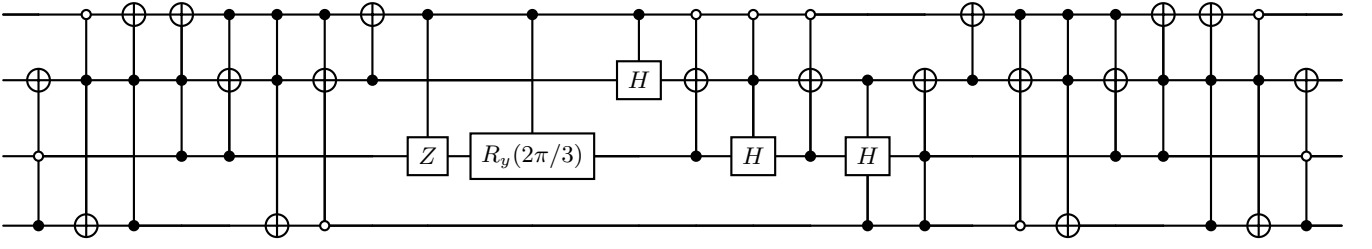

Unlike the generalized two-photon construction presented in Sec.\ref{sec:generalised_2_photon}, the three-photon scheme does not require an additional ancilla qubit. For any beam splitter acting between two spatial modes, the complete local Hilbert space up to three photons is contained within the four qubits assigned to those modes. The beam-splitter transformation can therefore be implemented directly as a four-qubit repeating unit, shown in Fig.~\ref{fig:four_qubit_circuits}. This unit is constructed from the circuit in Fig.\ref{fig:max_three_photon_circuit}, which is defined in the encoding of Table.\ref{tab:full_encoding}. Since the generalized (n)-mode construction instead employs the mode-wise Gray-code encoding of Table.\ref{tab:scalable_encoding} the Toffoli and CNOT gates at the beginning of the repeating unit convert the relevant four-qubit states from the Gray-code representation to the encoding of Table.\ref{tab:full_encoding}. The central part of the circuit then implements the corresponding beam-splitter transformation, after which the final Toffoli and CNOT gates restore the original Gray-code representation. In this way, the same four-qubit unit can be applied to any pair of modes in the interferometer without introducing additional ancilla qubits.

For a beam splitter acting between modes $i$ and $j$, the repeating unit acts on the qubits
\begin{equation}
(q_{2i-1},q_{2i},q_{2j-1},q_{2j}),
\end{equation}
while all remaining qubits remain unchanged. Therefore, each beam splitter in the optical interferometer can be replaced by its corresponding repeating unit. If the interferometer consists of the sequence
\begin{equation}
B_{i_1j_1},B_{i_2j_2},\ldots,B_{i_mj_m},
\end{equation}
the corresponding quantum circuit is
\begin{equation}
U_{\mathrm{QC}}^{(3)}
=
R^{(3)}_{i_mj_m}
\cdots
R^{(3)}_{i_2j_2}
R^{(3)}_{i_1j_1},
\end{equation}
where $R^{(3)}_{ij}$ denotes the repeating unit implementing the three-photon beam-splitter transformation between modes $i$ and $j$. By preserving the ordering and connectivity of the original interferometer, the constructed quantum circuit faithfully reproduces the corresponding three-photon boson-sampling evolution. Consequently, the proposed construction provides a scalable mapping from arbitrary $n$-mode linear-optical interferometers to gate-based quantum circuits using $2n$ qubits.

\subsection{Beam Splitter Transformation in the Single-Photon Sector}
\label{sec: BS_1_photon_transformation}

The action of a balanced beam splitter in the single-photon subspace can be represented by a two-qubit unitary acting on the computational basis
\(\{|00\rangle, |01\rangle, |10\rangle, |11\rangle\}\).
Here, the states \(|01\rangle\) and \(|10\rangle\) represent a single photon occupying the second and first spatial modes, respectively, while the states \(|00\rangle\) and \(|11\rangle\) remain unchanged. The transformation is given by
\begin{align}
|00\rangle &\rightarrow |00\rangle, \\
|01\rangle &\rightarrow \frac{|10\rangle-|01\rangle}{\sqrt{2}}, \\
|10\rangle &\rightarrow \frac{|10\rangle+|01\rangle}{\sqrt{2}}, \\
|11\rangle &\rightarrow |11\rangle.
\end{align}
The corresponding unitary matrix is
\begin{equation}
U_{\mathrm{BS}}=
\begin{pmatrix}
1 & 0 & 0 & 0\\
0 & -\frac{1}{\sqrt{2}} & \frac{1}{\sqrt{2}} & 0\\
0 & \frac{1}{\sqrt{2}} & \frac{1}{\sqrt{2}} & 0\\
0 & 0 & 0 & 1
\end{pmatrix},
\end{equation}
where each column corresponds to the transformed computational basis state. This unitary reproduces the beam-splitter action on the single-photon subspace while leaving the vacuum and auxiliary states invariant. Such an embedding is convenient for implementing beam-splitter operations within the standard gate-based quantum circuit formalism.

\subsection{Beam Splitter Transformation in the Two-Photon Sector}
\label{sec: BS_2_photon_transformation}

The action of a balanced beam splitter in the two-photon subspace can be represented by a two-qubit unitary acting on the computational basis
\(\{|00\rangle, |01\rangle, |10\rangle, |11\rangle\}\).
The encoding is chosen such that
\(|00\rangle \equiv |0,0\rangle\),
\(|01\rangle \equiv |0,2\rangle\),
\(|10\rangle \equiv |2,0\rangle\), and
\(|11\rangle \equiv |1,1\rangle\).
Under a balanced beam splitter, the basis states transform according to

\begin{align}
|00\rangle &\rightarrow |00\rangle, \\
|01\rangle &\rightarrow \frac{|10\rangle+|01\rangle}{2}
-\frac{|11\rangle}{\sqrt{2}}, \\
|10\rangle &\rightarrow \frac{|10\rangle+|01\rangle}{2}
+\frac{|11\rangle}{\sqrt{2}}, \\
|11\rangle &\rightarrow
\frac{|10\rangle-|01\rangle}{\sqrt{2}}.
\end{align}

The corresponding unitary matrix is

\begin{equation}
U_{\mathrm{BS}}=
\begin{pmatrix}
1 & 0 & 0 & 0\\
0 & \frac{1}{2} & \frac{1}{2} & -\frac{1}{\sqrt{2}}\\
0 & \frac{1}{2} & \frac{1}{2} & \frac{1}{\sqrt{2}}\\
0 & -\frac{1}{\sqrt{2}} & \frac{1}{\sqrt{2}} & 0
\end{pmatrix}.
\end{equation}

Each column of the matrix represents the transformed computational basis state. This unitary faithfully reproduces the two-photon beam-splitter dynamics, including the Hong--Ou--Mandel interference responsible for bosonic bunching. The embedding enables the beam-splitter transformation to be implemented using standard gate-based quantum circuits while preserving the underlying two-photon interference.

\subsection{Theoretical Transition Probability Matrix}

The four-mode linear optical interferometer considered in this work consists of three balanced beam splitters. Using the real Hadamard convention for a 50:50 beam splitter, the overall interferometer is described by the unitary matrix

\begin{equation}
U=
\begin{pmatrix}
\frac{1}{\sqrt2} & \frac{1}{\sqrt2} & 0 & 0\\
\frac12 & -\frac12 & \frac12 & \frac12\\
\frac12 & -\frac12 & -\frac12 & -\frac12\\
0 & 0 & \frac{1}{\sqrt2} & -\frac{1}{\sqrt2}
\end{pmatrix}.
\label{eq:interferometer}
\end{equation}

The theoretical transition probability between an input Fock state
$S=(s_1,s_2,s_3,s_4)$ and an output Fock state
$T=(t_1,t_2,t_3,t_4)$ is calculated using the boson sampling transition rule

\begin{equation}
P_{S\rightarrow T}
=
\frac{\left|\mathrm{Per}(U_{S,T})\right|^{2}}
{\prod_{i=1}^{4}s_i!\prod_{j=1}^{4}t_j!},
\label{eq:permanent_probability}
\end{equation}

where $U_{S,T}$ is the submatrix obtained by repeating the columns of
$U$ according to the input occupation numbers and the rows according to
the output occupation numbers. Here, $\mathrm{Per}(\cdot)$ denotes the
matrix permanent.

The two-photon Fock basis is encoded into the four-qubit computational
basis according to

\begin{equation}
\begin{aligned}
|2000\rangle &\rightarrow |1000\rangle, &
|0200\rangle &\rightarrow |0100\rangle,\\
|0020\rangle &\rightarrow |0010\rangle, &
|0002\rangle &\rightarrow |0001\rangle,\\
|1100\rangle &\rightarrow |1100\rangle, &
|1010\rangle &\rightarrow |1010\rangle,\\
|1001\rangle &\rightarrow |1001\rangle, &
|0110\rangle &\rightarrow |0110\rangle,\\
|0101\rangle &\rightarrow |0101\rangle, &
|0011\rangle &\rightarrow |0011\rangle.
\end{aligned}
\label{eq:encoding}
\end{equation}

Using the above ordering, the theoretical transition probability matrix,
$P_{\mathrm{th}}$ (Eq. \eqref{eq:probability_matrix}, is constructed by evaluating
Eq.~(\eqref{eq:permanent_probability}) for every pair of input and output
basis states. The columns of $P_{\mathrm{th}}$ correspond to the input
states, while the rows correspond to the output states. Thus, the matrix
element $(i,j)$ represents the probability of obtaining the $i^{\mathrm{th}}$
output state given the $j^{\mathrm{th}}$ input state. Consequently, each
column satisfies the normalization condition
\begin{equation}
\sum_{i=1}^{10} P_{\mathrm{th}}(i,j)=1,
\qquad
j=1,\ldots,10.
\end{equation}

The resulting theoretical transition probability matrix is
\begin{equation}
\resizebox{0.50\textwidth}{!}{$
P_{\mathrm{th}}=
\begin{pmatrix}
0.25 & 0.25 & 0    & 0    & 0.50 & 0    & 0    & 0    & 0    & 0    \\
0.0625 & 0.0625 & 0.0625 & 0.0625 & 0.125 & 0.125 & 0.125 & 0.125 & 0.125 & 0.125 \\
0.0625 & 0.0625 & 0.0625 & 0.0625 & 0.125 & 0.125 & 0.125 & 0.125 & 0.125 & 0.125 \\
0    & 0    & 0.25 & 0.25 & 0    & 0    & 0    & 0    & 0    & 0.50 \\
0.25 & 0.25 & 0    & 0    & 0    & 0.125 & 0.125 & 0.125 & 0.125 & 0    \\
0.25 & 0.25 & 0    & 0    & 0    & 0.125 & 0.125 & 0.125 & 0.125 & 0    \\
0    & 0    & 0    & 0    & 0    & 0.25 & 0.25 & 0.25 & 0.25 & 0    \\
0.125 & 0.125 & 0.125 & 0.125 & 0.25 & 0    & 0    & 0    & 0    & 0.25 \\
0    & 0    & 0.25 & 0.25 & 0    & 0.125 & 0.125 & 0.125 & 0.125 & 0    \\
0    & 0    & 0.25 & 0.25 & 0    & 0.125 & 0.125 & 0.125 & 0.125 & 0
\end{pmatrix}
$}
\label{eq:probability_matrix}
\end{equation}
\vspace{0.3cm}

This theoretical probability matrix serves as the reference for
comparison with the experimentally measured transition probability matrix
presented in Section~\ref{sec:results}.






\subsection{Generalized Framework for Arbitrary \texorpdfstring{$m$}{m}-Photon, \texorpdfstring{$n$}{n}-Mode Boson Sampling}

This appendix outlines a systematic procedure for extending the quantum circuit framework developed in this paper to arbitrary $m$-photon, $n$-mode boson sampling. The central idea is to first construct quantum circuits corresponding to the action of a single beam splitter on different photon-number sectors and then recursively combine these sectors into a unified beam-splitter circuit acting on the complete truncated Fock space. Finally, this beam-splitter unit is embedded into an arbitrary boson sampling interferometer through a scalable mode-to-qubit encoding.

\subsubsection*{Step 1: Construct Beam-Splitter Circuits for Individual Photon-Number Sectors}

The first step is to derive the quantum circuit corresponding to a single beam splitter for each possible input photon number. Specifically, construct the circuit for
\begin{enumerate}[label=(\alph*)]
    \item One photon incident on a beam splitter.
    \item Two photons incident on a beam splitter.
    \item Three photons incident on a beam splitter.
    \item $\vdots$
    \item $m$ photons incident on a beam splitter.
\end{enumerate}
The motivation for constructing each of these circuits individually is that, although an arbitrary boson sampling interferometer contains a total of $m$ photons, the number of photons entering any particular beam splitter is not fixed. Depending on the preceding transformations in the interferometer, a beam splitter may receive any number of photons ranging from $0$ to $m$.

Therefore, the beam-splitter circuit must correctly implement its transformation on every photon-number subspace,
\[
0,1,2,\ldots,m.
\]
Each photon-number sector has a different Hilbert-space dimension and therefore requires an independent qubit encoding and circuit construction.

\subsubsection*{Step 2: Recursively Combine the Photon-Number Sectors}

After constructing the beam-splitter circuits for the individual photon-number sectors, they are recursively combined into a single circuit acting on the complete truncated Fock space.

The construction begins by combining the vacuum, one-photon and two-photon subspaces into a single beam-splitter circuit, denoted by $X_1$. Since the two-photon transformation should only act on states belonging to the two-photon subspace, the corresponding gates are applied conditionally using controlled operations that distinguish this subspace from the vacuum and single-photon sectors.

The same procedure is then repeated recursively. The three-photon beam-splitter circuit is combined with $X_1$ to obtain a circuit acting on the combined Hilbert space
\[
0,1,2,3.
\]
The gates corresponding to the three-photon transformation are again activated only when the encoded state belongs to the three-photon subspace. This selective operation is achieved using multi-controlled gates, such as Toffoli gates, together with ancilla qubits whenever required.

The procedure is repeated successively by incorporating the four-photon, five-photon and higher photon-number beam-splitter circuits until a single circuit is obtained that correctly implements the beam-splitter transformation on the complete truncated Hilbert space,
\[
0,1,2,\ldots,m.
\]
An important feature of this recursive construction is the choice of encoding. As additional photon-number sectors are incorporated,
\begin{itemize}
    \item the size of the Fock basis increases,
    \item the number of required encoding qubits increases,
    \item additional ancilla qubits may be required for subspace identification, and
    \item the optimal encoding may change.
\end{itemize}
Consequently, the encoding employed during this recursive construction need not be scalable. Instead, at each stage an encoding should be chosen that minimizes circuit complexity while enabling efficient implementation of the required controlled operations.

Let the encoding used by the final beam-splitter circuit be denoted by
\[
M_1:\{r_1,r_2,r_3,\ldots\},
\]
where $\{r_i\}$ represents the encoded computational basis states. Let the resulting beam-splitter circuit be denoted by $C$.

\subsubsection*{Step 3: Embedding the Beam-Splitter Circuit into an Arbitrary Boson Sampling Network}

The encoding $M_1$ is selected for efficient beam-splitter implementation and therefore may not be suitable for representing large interferometers. To obtain a scalable framework, a separate scalable mode-to-qubit encoding is introduced.

Let
\[
M_2:\{q_1,q_2,\ldots,q_n\},
\]
denote a scalable mapping from the optical modes of the interferometer to qubits, where $n$ is determined by the chosen encoding and is generally greater than or equal to the number of modes.

Before applying the beam-splitter circuit $C$, the computational basis represented using $M_2$ is transformed into the internal encoding $M_1$. This basis conversion can be implemented using reversible logic composed of CNOT, Toffoli and other multi-controlled gates \cite{barenco1995elementary}.

The overall sequence for implementing each beam splitter is therefore
\[
M_2
\longrightarrow
M_1
\xrightarrow{\;C\;}
M_1
\longrightarrow
M_2.
\]
The first transformation converts the scalable encoding into the internal encoding required by the beam-splitter circuit. The circuit $C$ then performs the desired beam-splitter transformation, after which the inverse conversion restores the scalable encoding so that subsequent beam splitters can be implemented in the same manner.

This modular construction allows the same beam-splitter circuit to be reused throughout an arbitrary boson sampling interferometer while maintaining a scalable global encoding. Consequently, once the universal beam-splitter circuit has been constructed for photon numbers up to $m$, it can be repeatedly employed to realize quantum circuits corresponding to arbitrary $m$-photon, $n$-mode boson sampling networks. The ability to construct circuits in regimes where the number of
modes scales linearly with the photon number is particularly relevant
in light of recent complexity-theoretic results establishing hardness
evidence for BosonSampling in the saturated, linear-mode regime
\cite{bouland2026complexity}.


\end{document}